\documentclass[floatfix,twocolumn,amsmath,amssymb,superscriptaddress,aps,pre]{revtex4-2}
\usepackage{color}
\usepackage[usenames,dvipsnames,svgnames,table]{xcolor}
\usepackage[colorlinks=true,linkcolor=blue,urlcolor=blue,citecolor=blue]{hyperref}
\usepackage{mathtools}
\usepackage{graphicx}
\usepackage{dcolumn}
\usepackage{array}
\usepackage{lipsum}
\usepackage{bm}
\usepackage{subfigure}
\usepackage{amssymb}
\usepackage{multirow}
\usepackage{tabularx}
\usepackage{amsmath}
\usepackage{braket}
\usepackage{csquotes}
\graphicspath{{plots/}}
 \usepackage{lipsum}
\usepackage{mathrsfs}
\usepackage{MnSymbol}
\usepackage{textalpha}

\newcommand{\beq}{\begin{equation}}
\newcommand{\eeq}{\end{equation}}
\newcommand{\bea}{\begin{eqnarray}}
\newcommand{\eea}{\end{eqnarray}}

\begin{document}

\title{Spectral Deconvolution without the Deconvolution: Extracting Temperature from X-ray Thomson Scattering Spectra without the Source-and-Instrument Function}

\author{Thomas~Gawne}
\email{t.gawne@hzdr.de}
\affiliation{Center for Advanced Systems Understanding (CASUS), D-02826 G\"orlitz, Germany}
\affiliation{Helmholtz-Zentrum Dresden-Rossendorf (HZDR), D-01328 Dresden, Germany}

\author{Alina~Kononov}
\affiliation{Center for Computing Research, Sandia National Laboratories, Albuquerque NM 87185, USA}

\author{Andrew~D.~Baczewski}
\affiliation{Center for Computing Research, Sandia National Laboratories, Albuquerque NM 87185, USA}

\author{Hannah~M.~Bellenbaum}
\affiliation{Center for Advanced Systems Understanding (CASUS), D-02826 G\"orlitz, Germany}
\affiliation{Helmholtz-Zentrum Dresden-Rossendorf (HZDR), D-01328 Dresden, Germany}
\affiliation{Institut f\"ur Physik, Universit\"at Rostock, D-18057 Rostock, Germany}

\author{Maximilian~P.~B\"ohme}
\affiliation{Quantum Simulations Group, Physics and Life Science Directorate, Lawrence Livermore National Laboratory (LLNL), California 94550 Livermore, USA}

\author{Zhandos~A.~Moldabekov}
\affiliation{Institute of Radiation Physics, Helmholtz-Zentrum Dresden-Rossendorf (HZDR), D-01328 Dresden, Germany}

\author{Thomas~R.~Preston}
\affiliation{European XFEL, D-22869 Schenefeld, Germany}

\author{Sebastian~Schwalbe}
\affiliation{Center for Advanced Systems Understanding (CASUS), D-02826 G\"orlitz, Germany}
\affiliation{Helmholtz-Zentrum Dresden-Rossendorf (HZDR), D-01328 Dresden, Germany}

\author{Jan~Vorberger}
\affiliation{Institute of Radiation Physics, Helmholtz-Zentrum Dresden-Rossendorf (HZDR), D-01328 Dresden, Germany}

\author{Tobias~Dornheim}
\affiliation{Institute of Radiation Physics, Helmholtz-Zentrum Dresden-Rossendorf (HZDR), D-01328 Dresden, Germany}
\affiliation{Center for Advanced Systems Understanding (CASUS), D-02826 G\"orlitz, Germany}

\begin{abstract}
X-ray Thomson scattering (XRTS) probes the dynamic structure factor of the system, but the measured spectrum is broadened by the combined source-and-instrument function (SIF) of the setup. In order to extract properties such as temperature from an XRTS spectrum, the broadening by the SIF needs to be removed. Recent work [Dornheim~\textit{et al.} Nature Commun. \textbf{13}, 7911 (2022)] has suggested that the SIF may be deconvolved using the two-sided Laplace transform. However, the  extracted information can depend strongly on the shape of the input SIF, and the SIF is in practice challenging to measure accurately. Here, we propose an alternative approach: we demonstrate that considering ratios of Laplace-transformed XRTS spectra collected at different scattering angles is equivalent to performing the deconvolution, but without the need for explicit knowledge of the SIF. From these ratios, it is possible to directly extract the temperature from the scattering spectra, when the system is in thermal equilibrium. We find the method to be generally robust to spectral noise and physical differences between the spectrometers, and we explore situations in which the method breaks down. Furthermore, the fact that consistent temperatures can be extracted for systems in thermal equilibrium indicates that non-equilibrium effects could be identified by inconsistent temperatures of a few eV between the ratios of three or more scattering angles.
\end{abstract}

\maketitle

\section{Introduction\label{sec:introduction}}

High energy density (HED) systems are routinely produced in the laboratory at high energy laser facilities~\cite{hoarty2013observations,Fletcher_2014_Observations,Kraus_Science_2022,Tilo_Nature_2023} and x-ray free electron laser (XFEL) facilities~\cite{Vinko2012-fc,Ciricosta_2016_IPD,kraus_xrts,Kraus2025}, and are commonplace in astrophysical settings~\cite{guillot1999interiors,Bailey2015-hi,Kritcher2020}.
However, these systems are difficult to diagnose in the laboratory: due to their characteristically high temperatures and pressures, they are transient in the laboratory, and the breakdown of these conditions comes with the destruction of the sample. We therefore rely on a variety of \textit{in-situ} spectroscopy methods to diagnose systems in the short times they exist~\cite{Vinko2012-fc,Ciricosta_2016_IPD,Gorman_2024_Shock,Crepisson_2025_Shock,Harmand_2015_XAS,Harmand_2023_XAS,Humphries_2020_RIXS,Forte_2024_RIXS,siegfried_review,kraus_xrts,Dornheim_review}. 
At the same time, HED systems are very difficult to model owing to the complex interplay between thermal, Coulomb coupling, and quantum effects, which all need to be accounted for~\cite{wdm_book,review,H_Review,Vorberger_2025_Roadmap}. Well-characterised experiments are therefore vital for the benchmarking of theoretical models of such systems. 
However, very often the interpretation of measured spectra relies on the very models we need to benchmark. And, in many cases, the modelling of spectra is an ill-posed inverse problem, leading to a large range of possible conditions and theoretical models that can fit to the same spectrum~\cite{Kasim_PoP_2019,Hentschel_2025_Statistical}.

\begin{figure*}
    \centering
    \begin{subfigure}
        \centering
        \includegraphics[width=0.47\linewidth,keepaspectratio]{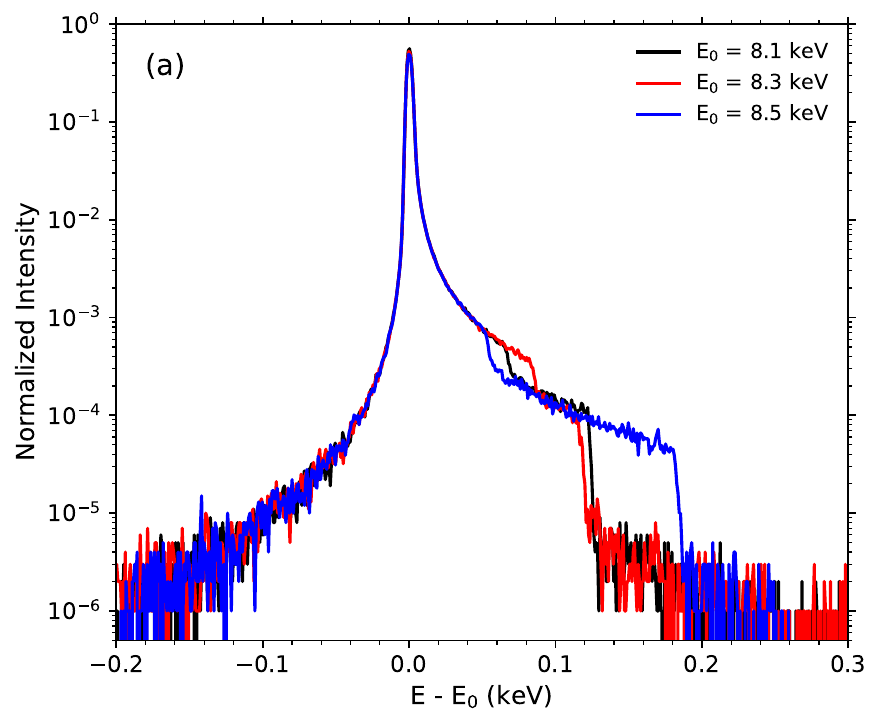}
    \end{subfigure}
    ~
    \begin{subfigure}
        \centering
        \includegraphics[width=0.47\linewidth,keepaspectratio]{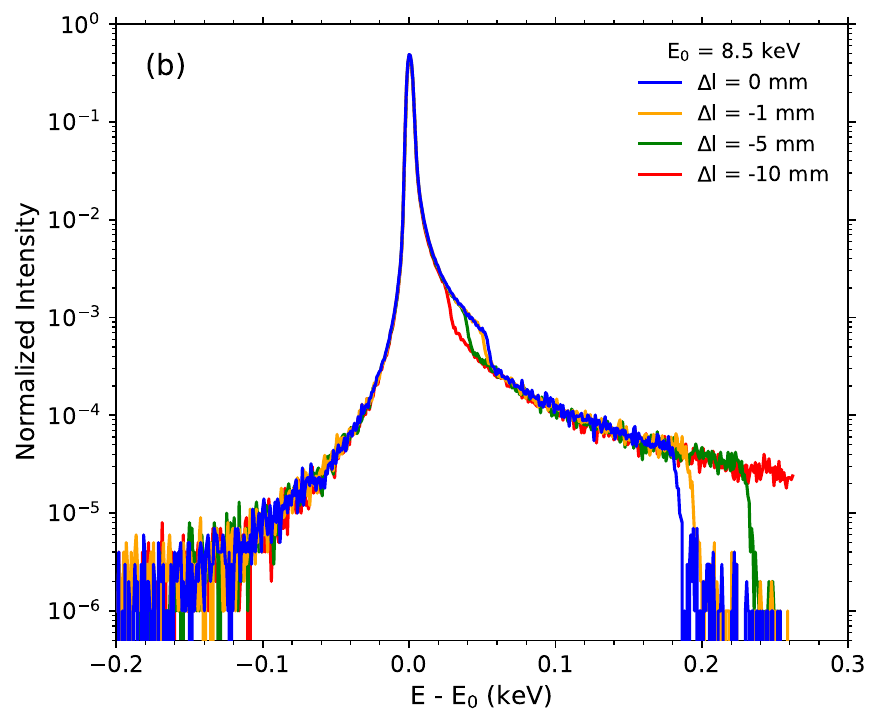}
    \end{subfigure}
    \caption{(a) Ray traced IFs on the spectrometer used in this work, which are representative of typical IFs for HAPG von H\'amos spectrometers. The IFs are measured with different single photon energies $E_0$ on the crystal: 8.1~keV (black), 8.3~keV (red), and 8.5~keV (blue). (b) Ray traced IFs at 8.5~keV with the crystal at different positions $\Delta l$ along the dispersion axis. In both plots, the sudden drops in the intensity correspond to the physical edges of a crystal in space being reached, so only part of the crystal reflects photons to the detector (see Ref.~\cite{Gawne_2024_Effects} for further details).
    }
    \label{fig:ExampleIFs}
\end{figure*}

X-ray Thomson scattering (XRTS) has emerged as a workhorse spectroscopy technique in the diagnosis of HED systems~\cite{Gregori_PRE_2003,Glenzer_2007_Observations,Kritcher_Science_2008,kraus_xrts,Tilo_Nature_2023,Sperling_PRL_2015,Fletcher_2014_Observations,Fletcher2015,Doeppner_2014,schwalbe2025staticlineardensityresponse,Dornheim_2024_Wr,Fletcher_JET_2022,Dornheim_2025_Unraveling,Boehme_2023_FreeBound}. By scattering photons of energy $E_i$ off the electrons in a system, the electronic dynamic structure factor (DSF) $S(\bm{q}, E_i-E_s)$ can be probed along a given scattering vector $\bm{q}$ and spectrally resolved in the scattered photon energy $E_s$. The scattering spectrum emerging from the target $I_s(\bm{q},E_s)$ is the DSF broadened by the incident beam profile $B(E_i)$,
\begin{equation}
    I_s(\bm{q}, E_s) = \int_{-\infty}^{\infty} dE_i \, B(E_i) \left(\frac{E_s}{E_i}\right)^2 S(\bm{q}, E_i-E_s) \, .
    \label{eq:PowerSpec}
\end{equation}
To spectrally resolve the energy distribution of the scattered photons, a spectrometer containing a dispersive crystal is typically used. The scattering of photons inside the crystal leads to further broadening, which is characterised by the instrument function (IF) $C(E-E_s; \bm{\Gamma}(E_s))$~\cite{Gawne_2024_Effects}, leading to a final measured spectrum:
\begin{equation}
    I(\bm{q}, E) = \int_{-\infty}^{\infty} dE_s \, C(E-E_s; \bm{\Gamma}(E_s)) I_s(\bm{q}, E_s) \, ,
    \label{eq:SpecResponse}
\end{equation}
where $E$ is the photon energy as inferred by the spectrometer calibration, and $\bm{\Gamma}(E_s)$ are the parameters that describe the broadening by the crystal. Typically, these parameters depend on the energy of the scattered photon~\cite{zachariasen1994theory,Gerlach_JAC_2015}. The IF also depends strongly on where the crystal is positioned in space~\cite{Gawne_2024_Effects,HEART}, as shown in Fig.~\ref{fig:ExampleIFs}. Evidently, the crystal IF does not broaden the spectrum as a convolution.
Nevertheless, in practice, the IF broadening is often treated as a simple convolution to the spectrum. The final measured spectrum is then just treated as the DSF convolved with the combined source-and-instrument function (SIF) $R(E)$ of the setup:
\begin{equation}
    I(\bm{q}, E) \approx \int_{-\infty}^{\infty} dE_i \, R(E_i) S(\bm{q}, E_i-E) \equiv (R \circledast S)(\bm{q}, E) \, ,
    \label{eq:XRTS_Convolution}
\end{equation}
where the approximation $E_s/E_i \approx 1$ across the spectral range has also been made.

The DSF contains a wealth of useful information on the temperature, density, ionization, and electronic correlations in the probed system~\cite{giuliani2008quantum,Crowley_2013}. But, in order to extract these properties from the DSF, the broadening by the SIF needs to be dealt with. Since direct deconvolution is usually unstable due to the spectral noise and finite spectral windows, the standard approach to extracting the DSF is so-called ``forward modelling'', where a model of the DSF is convolved with the SIF and fit to the spectrum; see e.g. Refs.~\cite{kraus_xrts,Fletcher_2014_Observations,Glenzer_2007_Observations,Kritcher_2011_In-flight,Boehme_2023_FreeBound}. The desired quantities -- e.g. temperature, density, and ionization -- are then the free fitting parameters. Consequently, the conditions extracted this way then depend directly on the assumptions made in the model (e.g. Refs.~\cite{kraus_xrts,Boehme_2023_FreeBound}), and therefore may not accurately reflect the true system conditions.

More recently~\cite{Dornheim_T_2022}, it has been proposed to examine the two-sided Laplace transformation of the DSF, the imaginary time correlation function (ITCF):
\begin{equation}\label{eq:define_ITCF}
    F(\mathbf{q},\tau) = \mathcal{L}\left[S(\mathbf{q}, \omega)\right](\tau) \equiv  \lim_{x\rightarrow\infty} \int_{-x}^x \textnormal{d}\omega\ S(\mathbf{q}, \omega)\ e^{-\tau \omega} \, ,
\end{equation}
where $\omega = E_i - E_s$ is the energy loss.
It is worth emphasising that since this is a one-to-one transformation, the ITCF contains exactly the same information as the DSF, just in a different representation~\cite{Dornheim_MRE_2023,Dornheim_PTR_2022,chuna2025secondrotonfeaturestrongly}.
For example, when a system is in thermal equilibrium, the detailed balance relationship of the DSF gives the temperature $k_\textnormal{B}T = \beta^{-1}$ of the system:
\begin{equation}\label{eq:DetailedBalance}
    S(\bm{q}, -\omega) = S(\bm{q}, \omega) e^{-\beta \omega } \, .
\end{equation}
In the ITCF, detailed balance is expressed in its symmetry around its minimum at $\tau = \beta/2$:
\begin{equation}
    F(\bm{q}, \tau) = F(\bm{q}, \beta - \tau) \, .
\end{equation}
If the system is not in equilibrium, then the ITCF is no longer symmetric, nor does it have the same minimum point in $\tau$ at different wavevectors $\bm{q}$, if it exhibits a minimum at all~\cite{Vorberger_Noneq,Bellenbaum_2025_Warm}.

The benefit of the ITCF method is that the two-sided Laplace transform has a convolution theorem and, since the kernel is real, the deconvolution is robust to spectral noise. Therefore, if the SIF is known and behaves sufficiently similarly to a convolution, the ITCF in Eq.~(\ref{eq:define_ITCF}) can be directly extracted from experimental spectra:
\begin{equation}\label{eq:convolution_theorem}
    F = \mathcal{L}\left[S\right] = \frac{\mathcal{L}\left[S\right] \mathcal{L}\left[R\right]}{\mathcal{L}\left[R\right]} = \frac{\mathcal{L}\left[S \circledast R\right]}{\mathcal{L}\left[R\right]} = \frac{\mathcal{L}\left[I\right]}{\mathcal{L}\left[R\right]} \, .
\end{equation}
While this transformation is only strictly valid when the integration boundaries in Eq.~(\ref{eq:define_ITCF}) become infinite (i.e. $x \rightarrow \infty$), in practice some properties, such as the symmetry point of the ITCF, converge within a finite $x$~\cite{Dornheim_T_2022, Dornheim_T2_2022}. The temperature can therefore be extracted by observing the convergence of the minimum of the ITCF with respect to increasing $x$.
The ITCF approach thus provides a way to directly extract temperature (among other physical properties) from an XRTS spectrum, without the need for a model or simulation result of the DSF.

One of the main practical challenges in using the ITCF method is the accurate characterization of the SIF~\cite{MacDonald_PoP_2021,Gawne_2024_Effects}.
Much of the difficulty is in characterising the IF of the mosaic crystals that are widely employed in XRTS experiments. Recent works have shown their IFs can extend over the whole detected spectral range with features that range over several orders of magnitude in intensity~\cite{Gawne_2024_Effects, HEART}; see Fig.~\ref{fig:ExampleIFs}. Due to the strong geometry dependence of the IF -- shown in particular by Fig.~\ref{fig:ExampleIFs}~(b) -- it is necessary to measure the SIF in every experiment.

One common approach for measuring the SIF directly is to measure the shape of the quasi-elastic scattering since, except for ultrahigh resolution setups~\cite{McBride_RSI_2018,Wollenweber_RSI_2021,Descamps_JSR_2022}, it just reproduces the SIF. However, eventually the quasi-elastic scattering will mix with inelastic scattering, meaning full characterization of the SIF is not possible.
In the case of x-ray backlighters~\cite{Park_2006_High,Tilo_Nature_2023,Doeppner_2014_MACS,Fletcher_2014_Observations,Kritcher_2011_In-flight}, there is also a strong dependency of the emission line intensities on the laser heating conditions for each shot. But, due to restrictions in the experimental setups, the actual emission spectrum for a given shot is rarely measured.
Therefore, models or approximations of the SIF are used in place of the actual SIF, which reintroduces a model-dependency in the underlying model-free analysis.

Here, we propose an alternative approach that avoids explicitly deconvolving the SIF. In XRTS experiments, it is very typical to collect scattering spectra from at least two scattering angles simultaneously in order to probe different regimes (e.g. collective versus non-collective). Returning to the definition of the convolution theorem in Eq.~(\ref{eq:convolution_theorem}), and presuming the SIF is the same between two spectrometers, it is easy to see that the ratio of the Laplace transforms of two spectra is equivalent to the ratios of their ITCFs:
\begin{equation}\label{eq:Ratios}
    \frac{\mathcal{L}\left[I(\bm{q}_1, E)\right]}{\mathcal{L}\left[I(\bm{q}_2, E)\right]} = \frac{\mathcal{L}\left[S(\bm{q}_1,  \omega)\right] \mathcal{L}\left[R(E)\right] }{\mathcal{L}\left[S(\bm{q}_2, \omega)\right] \mathcal{L}\left[R(E)\right]} = \frac{F(\bm{q}_1, \tau)}{F(\bm{q}_2, \tau)} \, .
\end{equation}
If the system is in thermal equilibrium, the two ITCFs are symmetric around the same minimum point, and therefore the ratio of the ITCFs will also be symmetric around this point. For the ratios, this may be a minimum or a maximum depending on the relative size and curvature of each ITCF, so for the purposes of this work we consider the minima of the indicated ratios. This means the temperature can be extracted directly from this ratio. With three or more scattering angles, comparisons of the ratios allows a consistency check: if the temperatures inferred are different between the ratios, or if the ratios themselves are not symmetric around their minimum or maximum, these would be key indicators of the system being out of equilibrium~\cite{Vorberger_Noneq}.
Crucially, by bypassing the explicit deconvolution of the SIF, this ratio method is truly model-free since no model of the DSF nor the SIF is required.

For the remainder of this article, we investigate the practicalities of the ratio method for temperature extraction in experiments. We perform ray tracing simulations of theoretical DSFs to test the ability of the method to reproduce known results on realistic spectrometers. As part of these investigations, we also consider the possibility of identifying non-equilibrium in the system via mismatches in the temperatures extracted between different ratios.
Additionally, we test the robustness of the method to spectral noise, the photon statistics, and variations in the IF between spectrometers.

\section{Simulation Details}\label{sec:sims}

\subsection{Dynamic Structure Factors}
Dynamic structure factors were calculated using the multi-component scattering simulation (MCSS) code~\cite{chapman2015probing,mcss_manual}, which uses the Chihara decomposition~\cite{Chihara_JoP_1987,Chihara_JoP_2000} to calculate the DSF. We simulated carbon with mass density 3.5~g~cm$^{-3}$ and a single C$^{4+}$ ion. The ions and electrons are in equilibrium with one another at temperatures of 15, 20, 50, and 100~eV.
The free-free feature was calculated using the random phase approximation (RPA)~\cite{quantum_theory,review}, while the bound-free and free-bound features were calculated with the impulse approximation (IA)~\cite{Eisenberger_1970_Compton,Mattern_2013_Theoretical}.
At the resolution of the spectrometers considered here, the details of the ion feature cannot be resolved, so the elastic feature is treated as a delta-function in frequency space multiplied by the Rayleigh weight. The Rayleigh weight then just determines the portion of the photons which elastically scatter. The screening cloud within the Rayleigh weight is calculated using finite-wavelength screening and a Coulomb potential with the effective charge of $Z=4$ for the electron-ion potential. The ionic form factor uses the simple hydrogenic orbital model for its description. The static ion structure is calculated by a hyper-netted chain solver using a Debye-H\"uckel potential featuring a general screening length for the ion-ion interaction.
The DSFs were calculated for a photon energy of 8.5~keV, and at scattering angles of 10$^\circ$, 20$^\circ$, and 150$^\circ$. We consider the spectrum incident on the spectrometer come from a single scattering vector, which is a reasonable approximation given the large incident photon energy compared to the size of the spectral window used in the deconvolution ($\le250$~eV)~\cite{Dornheim_T2_2022}.

\subsection{Ray Tracing}

\begin{figure}
    \centering
    \includegraphics[width=0.94\linewidth]{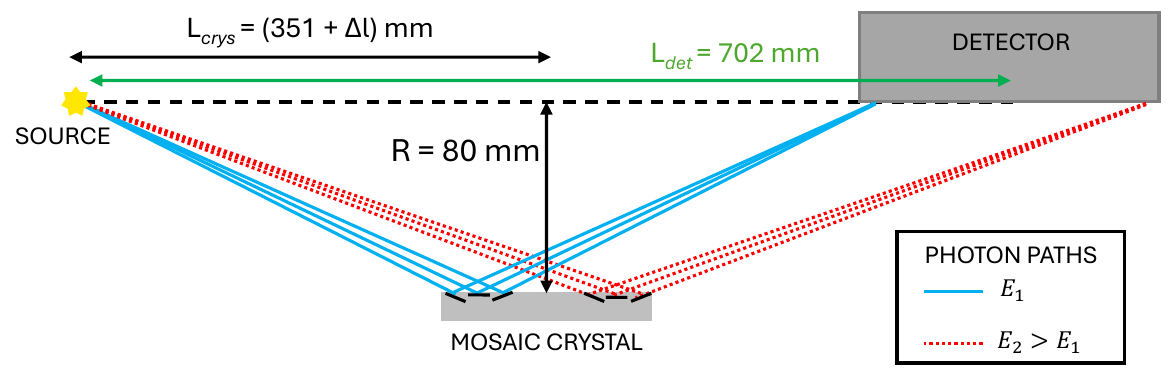}
    \caption{A two-dimensional sketch of the von H\'amos spectrometer setup used in this work. The crystal is centred at approximately 351~mm and positioned below the dispersion axis at the crystal's radius of curvature, 80~mm. The detector is fixed at 702~mm away from the source and sits in the dispersion plane. This setup corresponds to a central photon energy of 8.31~keV on the crystal and detector. In Section~\ref{sec:Alignment}, the crystal is moved along the dispersion axis by $\Delta l$ as well, but the detector remains fixed to maintain the spectral range.}
    \label{fig:SpecSetup}
\end{figure}

\begin{figure*}
    \centering
    \begin{subfigure}
        \centering
        \includegraphics[width=0.47\linewidth,keepaspectratio]{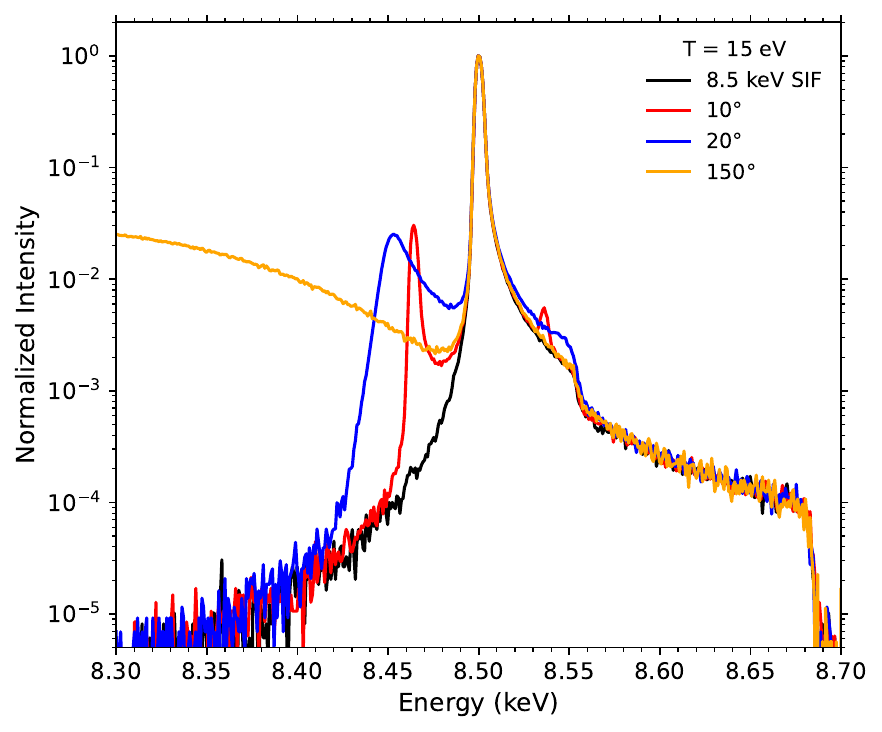}
    \end{subfigure}
    ~
    \begin{subfigure}
        \centering
        \includegraphics[width=0.47\linewidth,keepaspectratio]{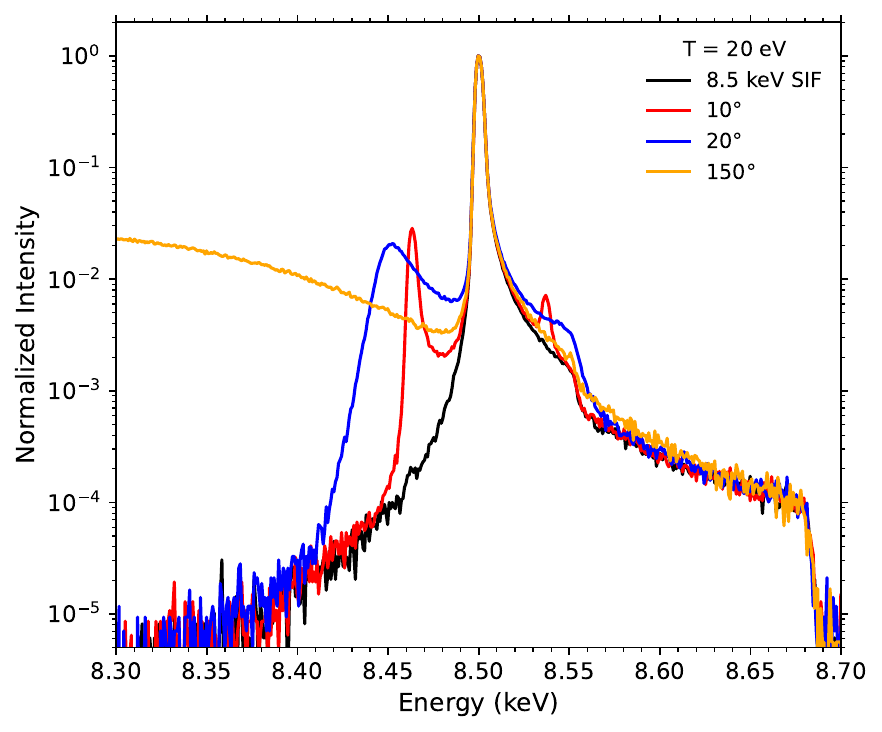}
    \end{subfigure}
    \vspace{0.1mm}
    \begin{subfigure}
        \centering
        \includegraphics[width=0.47\linewidth,keepaspectratio]{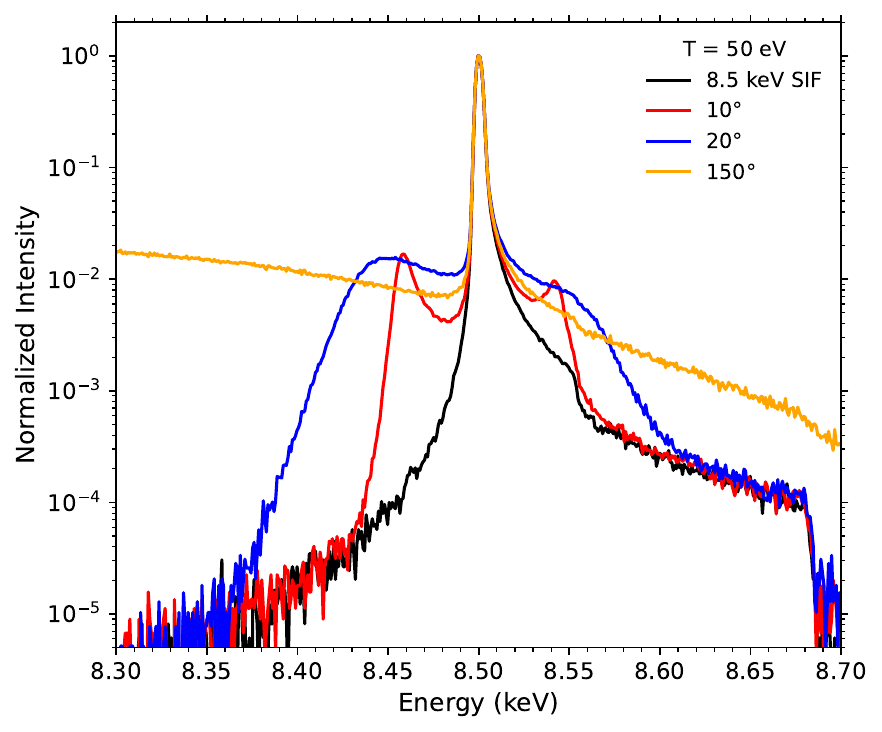}
    \end{subfigure}
    ~
    \begin{subfigure}
        \centering
        \includegraphics[width=0.47\linewidth,keepaspectratio]{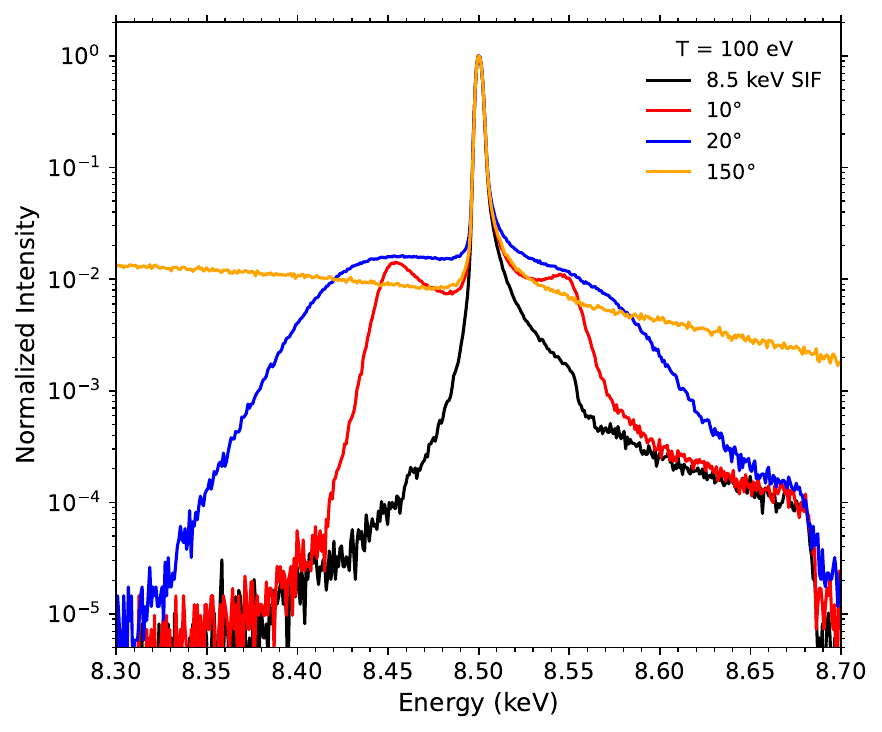}
    \end{subfigure}
    
    \caption{Ray traced XRTS spectra at different temperatures and scattering angles: 10$^\circ$ (red), 20$^\circ$ (blue), and 150$^\circ$ (orange). Also plotted is the SIF computed at 8.5~keV (black).}
    \label{fig:Spectra}
\end{figure*}

Ray tracing simulations were performed using HEART~v2~\cite{HEART,HEART_Gitlab}, a Monte Carlo ray tracer with specific support for mosaic crystals. We modelled a spectrometer composed of a highly annealed pyrolytic graphite (HAPG) crystal in  von H\'amos geometry~\cite{2009_Legall_Efficient,Zastrau_JoI_2013,Gerlach_JAC_2015}, identical to those used~\cite{Preston_JoI_2020} at the HED Scientific Instrument at the European XFEL~\cite{Zastrau_2021}: specifically, the crystal is 80~mm long, 30~mm wide, and 40~$\mu$m thick, with a radius of curvature of 80~mm, and operated in the (002) reflection. The detector has the dimensions of a Jungfrau detector~\cite{Mozzanica2016}, with $75\times75$~$\mu$m$^2$ pixels and dimensions $1024\times512$~pixels. The quantum efficiency of the detector is not treated, but it is relatively uniform in the spectral range considered here, so its effect would be rather minor, especially compared to the crystal's response function.
The crystal and detector are centred at 8.31~keV, providing a 250~eV spectral range above the elastic peak, which is quite typical in XRTS experiments. A two dimensional sketch of the setup is shown in Fig.~\ref{fig:SpecSetup}.

Based on fits to the quasi-elastic feature of polypropylene, the mosaic distribution of the crystallites was found to be well-described by a wrapped-Lorentzian-like distribution, with one of the crystals having a mosaic spread of 0.063$^\circ$~\cite{HEART}. Previous measurements of the mosaic spread in HAPG suggest there is some dependence of the mosaic spread on the photon energy below 11~keV~\cite{Gerlach_JAC_2015}, however the dependence is very weak, and so here it is assumed to be constant for all photon energies.

The intrinsic rocking curve (IRC) of the crystallites is modelled with a Voigt profile. For one of the HAPG crystals at the European XFEL, at 8.5~keV the Gaussian full width at half maximum (FWHM) is 50.0~$\mu$rad and the Lorentzian FWHM is 8.0~$\mu$rad~\cite{HEART}. The width and the peak of the IRC does show stronger dependence on photon energy than the mosaicity~\cite{zachariasen1994theory,Gerlach_JAC_2015}, so we account for it here. Specifically, we calculated intrinsic rocking curves of the crystallites~\cite{zachariasen1994theory} for a given distribution of their thicknesses, and fit the maxima and the FWHMs to Eqs.~(\ref{eq:IRC_FWHM})--(\ref{eq:IRC_FWHM2}). In doing this, we found the specific FWHM at a given photon energy strongly depends on the thickness distribution used, but the scaling between photon energies is relatively consistent between distributions.
Therefore, after selecting a given distribution, and calculating the FWHM and maximum positions, the FWHMs were rescaled to reproduce the total FWHM at 8.5~keV. The FWHM and maximum positions were then used to infer the following dependencies on photon energy:
\begin{align}
    W(E) &= 912.444[\mu \mathrm{rad~keV}] \frac{hc}{2d} \frac{1}{E^2} \frac{1}{\sin(2 \arcsin\left[\frac{hc}{2d} \frac{1}{E}\right]))}  \, , \label{eq:IRC_FWHM} \\
    M(E) &= 510.362[\mu \mathrm{rad~keV}] \frac{hc}{2d} \frac{1}{E^2} \frac{1}{\sin(2 \arcsin\left[\frac{hc}{2d} \frac{1}{E}\right]))} \, ,
    \label{eq:IRC_FWHM2}
\end{align}
where $W(E)$ is the FWHM function, $M(E)$ is the maximum point function, $h$ is Planck's constant, $c$ is the speed of light, and $d$ is the lattice spacing for the given Miller indices.
HEART also accounts for the dependency of the reflectivity and absorption in the crystal on photon energy~\cite{zachariasen1994theory}.
Taken together, this captures the complex dependency of the crystal rocking curve on the photon energy, on top of the geometric dependencies that can also be represented in terms of the photon energy.
This framework therefore constitutes the most advanced model for the influence of the instrument function, and far better than a simple convolution approach for exploring the applicability of the ratio method to real experiments.

Since the IF dominates the complexity of the full SIF, we treat the probing beam as a Gaussian centred at 8.5~keV with a FWHM of 1~eV. This is similar to having the beam pass through a monochromator. The initial distribution of photons that are ray traced is given by Eq.~(\ref{eq:PowerSpec}). The broadening of the spectra is dominated by the crystal, which substantially alters the shape and deconvolution behaviour in the ratio method. Still, monochromators see frequent use at XFELs to improve the spectral resolution of the measured DSF~\cite{Voigt_Ramen_2021,gawne2024ultrahigh,Gawne_2025_Si}, so our choice of source function is representative of realistic spectra collected at XFELs.

To produce the spectra, 10$^9$ photons were randomly sampled from the scattering spectrum. They were emitted from a point source, and uniformly distributed in the azimuthal and polar directions in a range that fully covers the crystal. This also means solid angle effects are accounted for since parts of the crystal further from the source have a lower angular coverage. The rays are then traced for their full path, until they either leave the crystal (i.e. multiple reflections are explicitly calculated) or the photon is absorbed. 

The simulation setup described above is most relevant for XFEL experiments. To demonstrate the applicability of the method to a broader range of experiments, in Section~\ref{sec:NIF} we consider a NIF-style experiment based on Ref.~\cite{Tilo_Nature_2023}. Further details on those simulations may be found in that section.

The detector images from the ray tracing simulations are output in terms of number of photons per pixel, so the resulting spectra consider number of detected photons per energy bin. While HEART is able to model filter and the quantum efficiency effects, usually these are either accounted for when processing a spectrum, or they are sufficiently uniform in the spectral range to be ignored. Additionally, for the detectors at XFELs, their response is typically calibrated ahead of experiments with known sources, allowing for a conversion of the analogue-to-digital units into the total photon energy deposited in a pixel. This can then be converted to photon counts via the spectrometer dispersion. Provided the spectrum units are proportional to photon counts per energy bin, the specific units are otherwise arbitrary since this just gives a constant in front of the ITCF. The ITCF could be normalised to its true units~\cite{dornheim2024fsum}, but this is unnecessary for the anaylsis performed here.

Examples of the ray traced spectra are shown in Fig.~\ref{fig:Spectra}.
Around 8.55~keV and 8.68~keV, there are sudden drops in the intensity of the elastic signal corresponding to the physical edges of the crystal in space (see Ref.~\cite{Gawne_2024_Effects} for further details). The shape of the elastic feature has a noticeable effect on the shape of the spectra for the low scattering angles and the low temperatures, where features are narrowest.
At high scattering angles for $T=50$ and 100~eV, the effect of the shape of the SIF is hidden by the very broad inelastic features.
Taken together, the extended nature of this SIF and its complexity can be challenging to measure, and deconvolving it from XRTS spectra effectively could be challenging.

The Laplace transform expressed in Eq.~(\ref{eq:define_ITCF}) is centred symmetrically around $\omega = 0$, which means a central photon energy needs to be defined. The choice is important since if an error $\Delta E$ is made, an additional factor of $\exp(-\tau \Delta E)$ appears in front of the ITCF which will skew the position of its minimum. Even though the central beam energy is 8.5~keV, this is not the appropriate choice due to the asymmetric nature of the SIF. Instead, the SIF intensity-weighted average of the spectrometer energy is used to define the central energy of the scattering spectrum.

Finally, the convergence of the inferred temperatures from the figures is identified visually, given the large variety in shape and form of the curves.

\section{Results\label{sec:results}}

\subsection{Checking the Convolution Theorem Approximation}
\begin{figure*}
    \centering
    \begin{subfigure}
        \centering
        \includegraphics[width=0.47\linewidth,keepaspectratio]{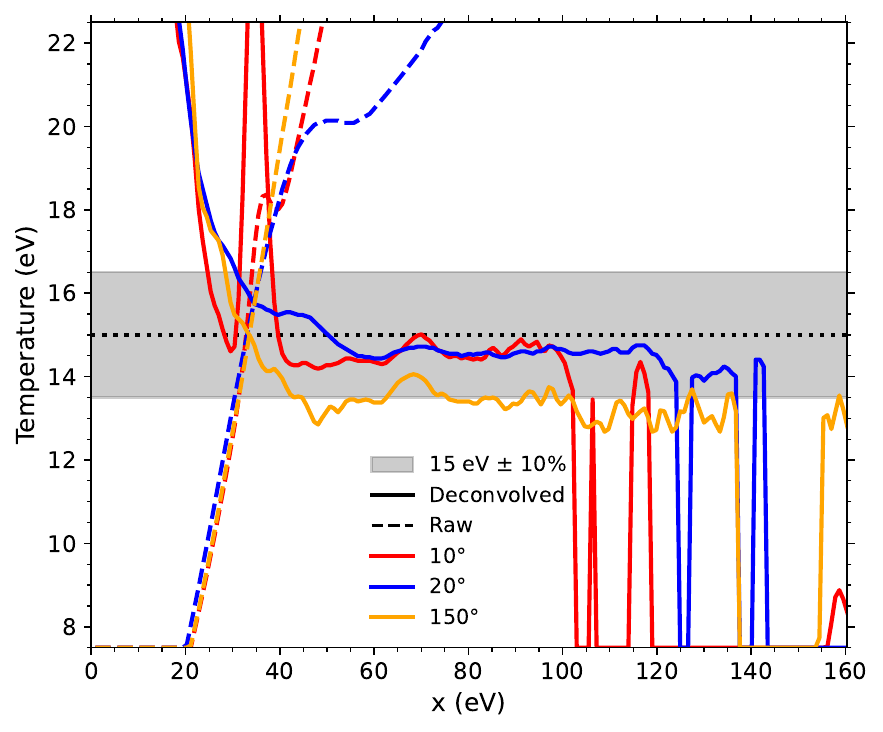}
    \end{subfigure}
    ~
    \begin{subfigure}
        \centering
        \includegraphics[width=0.47\linewidth,keepaspectratio]{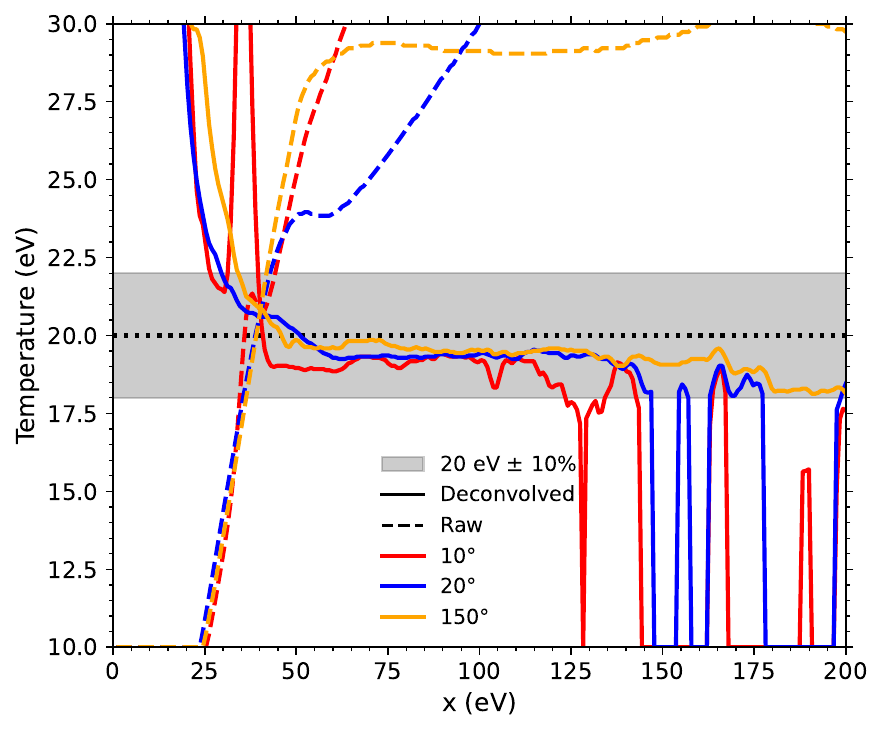}
    \end{subfigure}
    \vspace{0.1mm}
    \begin{subfigure}
        \centering
        \includegraphics[width=0.47\linewidth,keepaspectratio]{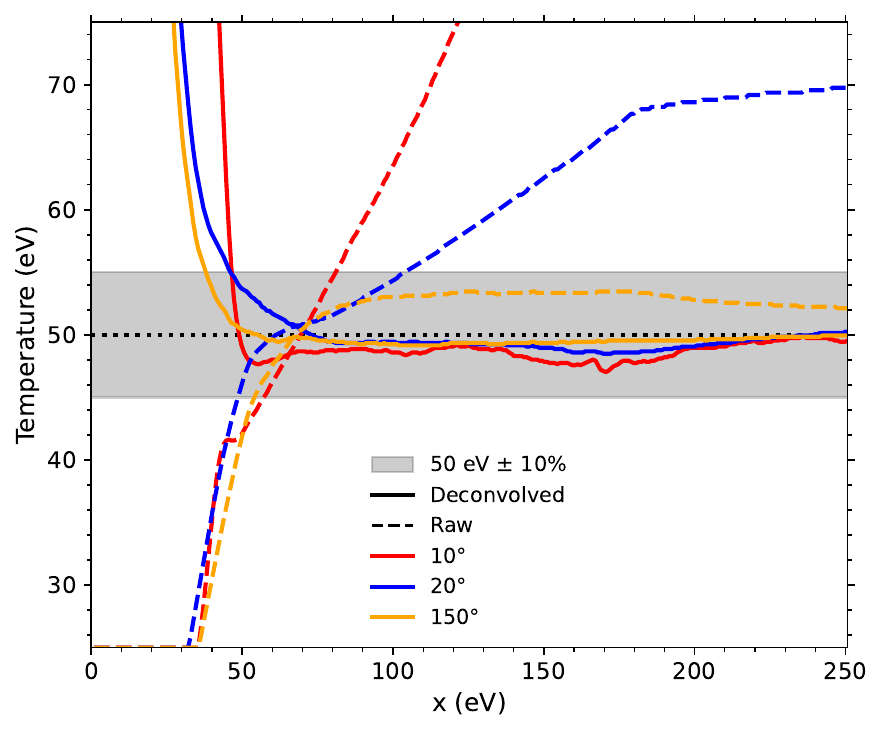}
    \end{subfigure}
    ~
    \begin{subfigure}
        \centering
        \includegraphics[width=0.47\linewidth,keepaspectratio]{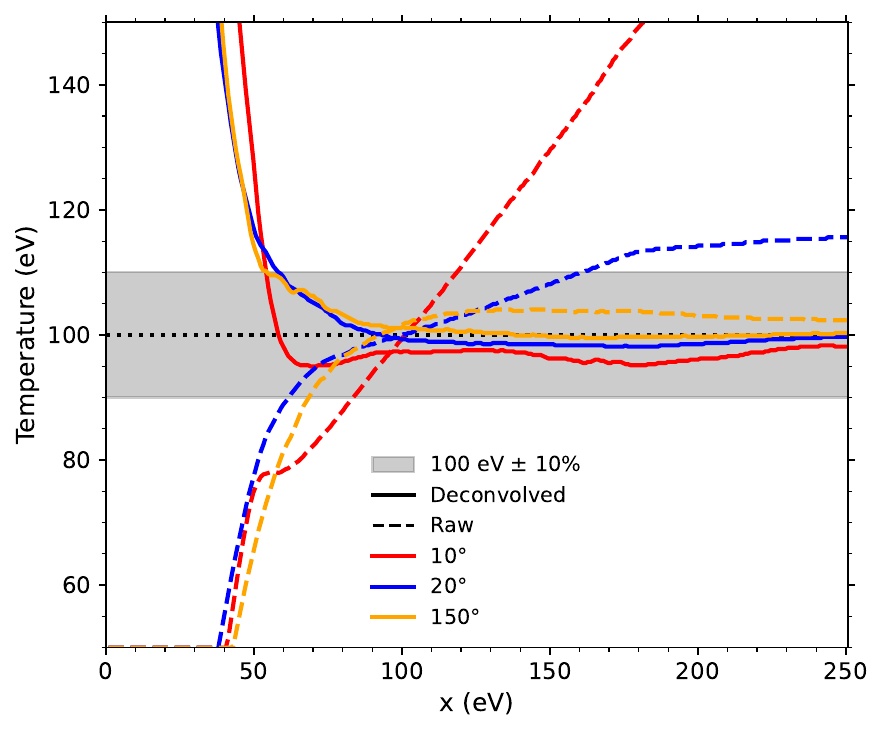}
    \end{subfigure}
    
    \caption{Convergence tests of the temperature extracted from the ray traced spectra in Fig.~\ref{fig:Spectra} via deconvolution of the SIF measured at 8.5~keV (solid lines) for different temperatures (15, 20, 50, 100~eV) and different scattering angles (10$^\circ$ in red, 20$^\circ$ in blue, and 150$^\circ$ in orange), with respect to the integration range $x$ in Eq.~(\ref{eq:define_ITCF}). The dashed lines show the minima of the Laplace transformed spectra, without removing the SIF. The dotted horizontal line indicates the exact temperature of the system, and the grey area indicates a $\pm10$\% interval around these temperatures.
    }
    \label{fig:DirectDeconv}
\end{figure*}

\begin{figure*}
    \centering
    \begin{subfigure}
        \centering
        \includegraphics[width=0.47\linewidth,keepaspectratio]{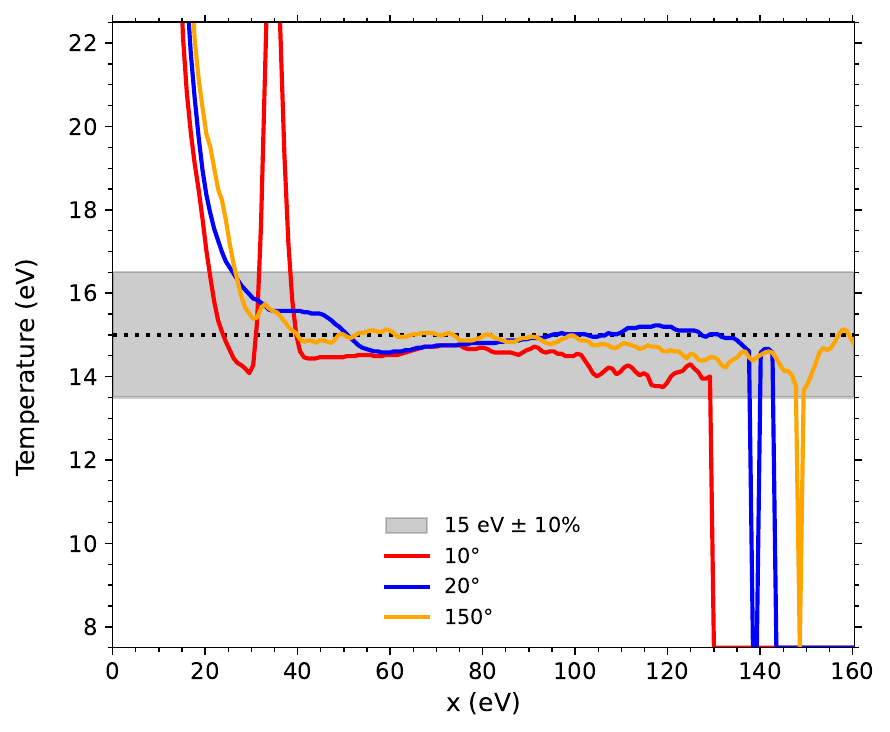}
    \end{subfigure}
    ~
    \begin{subfigure}
        \centering
        \includegraphics[width=0.47\linewidth,keepaspectratio]{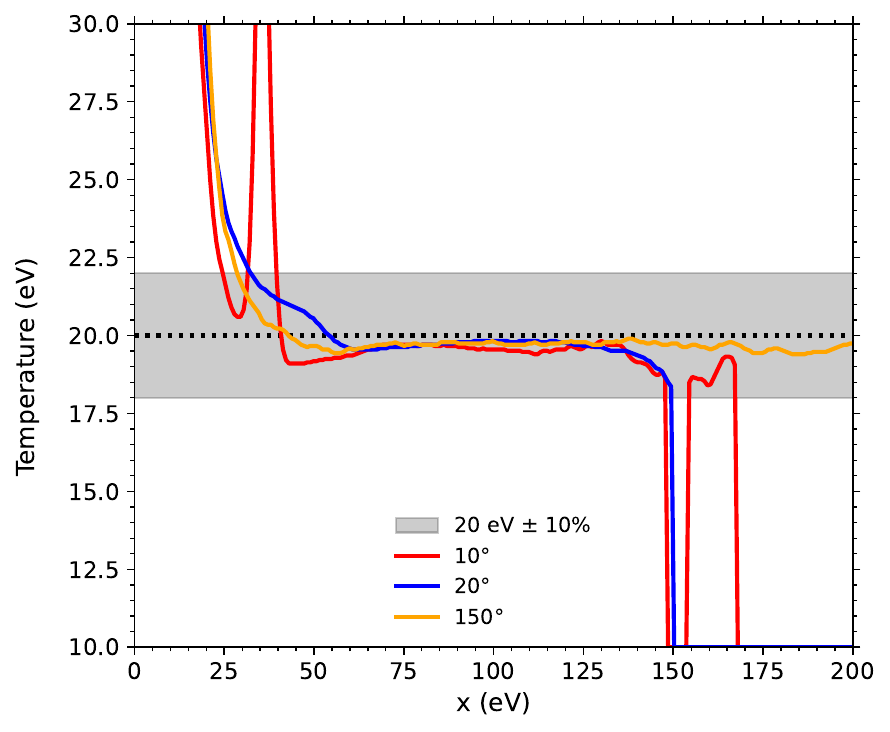}
    \end{subfigure}
    
    \caption{Convergence tests of the temperature extracted from the ray traced spectra for $T=15$~eV and 20~eV at the different scattering angles (10$^\circ$ in red, 20$^\circ$ in blue, and 150$^\circ$ in orange). Only the deconvolved curves are shown here. Compared to Fig.~\ref{fig:DirectDeconv}, the Rayleigh weight has been reduced by a factor of 4 here, so more photons are scattered inelastically.
    }
    \label{fig:ReducedElastic}
\end{figure*}

As mentioned previously, the ITCF method relies on the approximation that the SIF is applied as a convolution~\cite{Dornheim_T_2022,Dornheim_T2_2022}. However, this is not strictly true in the case of the crystal IF, and this approximation has yet to be confirmed. Given the ratio method proposed here also relies on this approximation, we first test whether the temperature can be extracted using the convolution theorem.

Fig.~\ref{fig:DirectDeconv} shows the convergences in the ITCF-inferred temperatures for the different spectra in Fig.~\ref{fig:Spectra}, with respect to the integration range $x$ in Eq.~(\ref{eq:define_ITCF}), when using the 8.5~keV SIF. Also shown are the minima of the Laplace transformed spectra without removing the SIF -- in general, these show no convergence to a specific temperature, highlighting the importance of removing the SIF broadening from the spectra.

In all cases, once the SIF is deconvolved from the spectrum, a region of convergence in the inferred temperature with respect to $x$ is observed. Oscillations in the convergence may be taken as an uncertainty in the inferred temperature. For all cases considered here, the temperature converges to the expected value within $\lesssim 10$\%.
The inferred temperature for 150$^\circ$ at 15~eV is on the -10\% limit, and this underestimation is likely due to the very weak upshifted inelastic signal for this spectrum (see Fig.~\ref{fig:Spectra}), so it is difficult to probe detailed balance. This still only represents an absolute error of 1.5~eV, which is comparable to the absolute error observed at the other temperatures.

In general, there is a small but seemingly systematic underestimation of the temperatures extracted here. This is likely due to the solid angle coverage of the crystal, which is larger on the parts of the crystal closer to the source, which corresponds to lower photon energies. Therefore, the spectrometer has a higher collection efficiency at lower photon energies. This means that it appears as if there are more downshifted photons than there are, which skews the apparent detailed balance. At the same time, the $(E_s/E_i)^2$ factor in Eq.~(\ref{eq:PowerSpec}) term mostly counteracts this effect by increasing the number of upshifted photons than detailed balance otherwise implies. While these effects will not be explicitly handled here since they mostly cancel, future improvements in analysis could account for them.

Otherwise, the stability of the convergences and accuracy of the inferred temperature demonstrates the ITCF method can accurately extract system conditions.
To within at least $\pm 250$~eV around the elastic feature, the convolution approximation appears to hold very well.

\begin{figure}
    \centering
    \includegraphics[width=\linewidth,keepaspectratio]{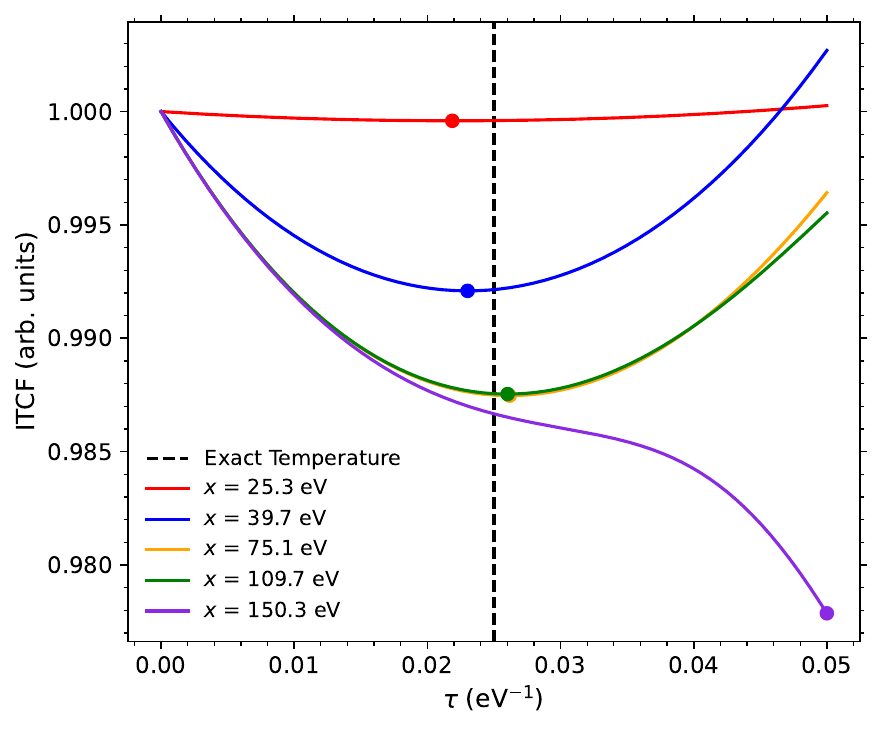}
    \caption{Examples of the ITCF for $T=20$~eV at scattering angle $10^\circ$ for different integration limits $x$. The location of the minima of each curve, which is used to infer temperature, are indicated by the circles. The minimum point of the exact temperature is indicated by the black dashed line.
    }
    \label{fig:ITCF_Examples}
\end{figure}

For the $T=15$~eV and 20~eV cases, sharp drops of the inferred temperature to $T_{\rm exact}/2$ are eventually seen as $x$ increases -- such behaviour will also be seen later when using the ratios of ITCFs. These correspond to the ITCF losing its minimum point as the upshifted signal becomes very weak and few, if any, photons are measured. This causes the $\tau>\beta/2$ side of the ITCF to collapse, at which point the ITCF minimum just becomes the upper limit of the considered $\tau$-range, here restricted to $\tau_{\rm max}\ = \beta$.
An example of this behaviour is shown for $T = 20$~eV at scattering angle $10^\circ$ in Fig.~\ref{fig:ITCF_Examples}, which plots the ITCFs for different integration limits. 
Such drops are not seen for $T=50$~eV and 100~eV since there is strong upshifted signal up to the edge of the detector.

These drops in the inferred temperature indicate the limits of being able to perform the deconvolution and extract temperature due to lack of upshifted signal. It also demonstrates that the lower temperature limit that the ITCF method can measure is ultimately determined by the photon statistics: enough signal needs to be measured sufficiently far from the elastic feature in both the upshifted and downshifted sides of the spectrum so that convergence in $x$ can be observed. The more extended the SIF is, the further away this point will be due to the need to deconvolve a broader function. And at the same time, the detailed balance relationship means the number of upshifted photons is exponentially reduced as the temperature decreases.

Likewise, the ratio of inelastic to elastic scattering matters since it is the inelastic signal which needs to be measured in order to probe detailed balance. To emphasise this point, Fig.~\ref{fig:ReducedElastic} shows the convergence of the inferred temperature for $T=15$~eV and 20~eV when the Rayleigh weight has been reduced by a factor of 4. In these cases, it takes a larger integration range $x$ before the ITCF loses its minimum, and the agreement between the temperatures at the different angles is improved. Convergences for a given number of input photons therefore have a strong dependence on the Rayleigh weight, and the models in MCSS are known to often overestimate the elastic scattering contribution~\cite{chapman2015probing,mcss_manual}. In the remainder of the article, the standard DSF model is used for the simulations, but will be compared to results for when the Rayleigh weight is reduced as well to understand the role of the elastic feature in extracting the temperature.

\subsection{Temperature from ITCF Ratios}\label{sec:NoSIFDeconv}
\begin{figure*}
    \centering
    \begin{subfigure}
        \centering
        \includegraphics[width=0.47\linewidth,keepaspectratio]{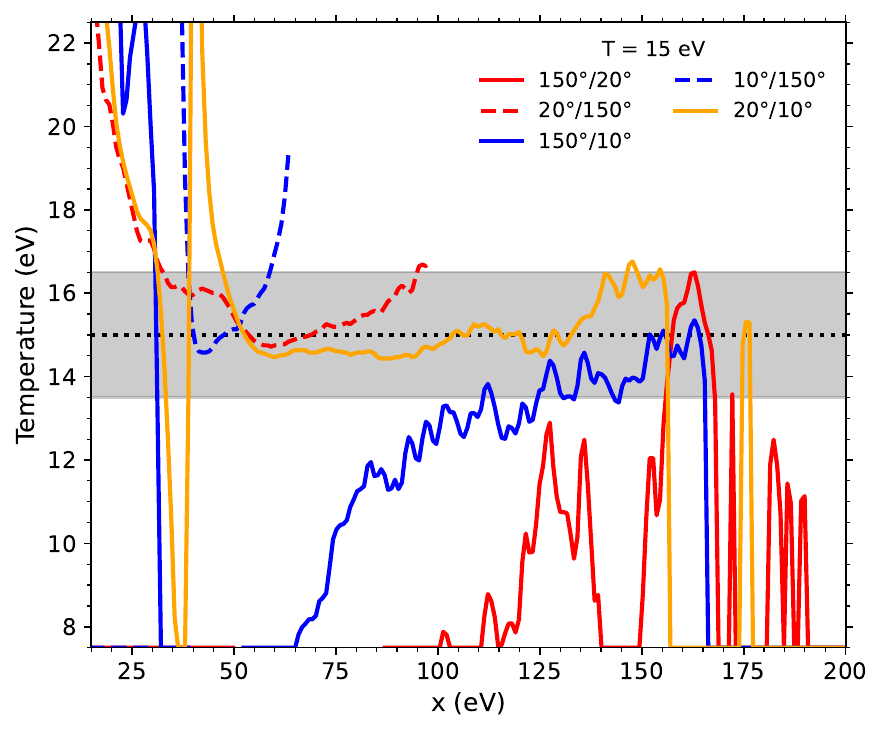}
    \end{subfigure}
    ~
    \begin{subfigure}
        \centering
        \includegraphics[width=0.47\linewidth,keepaspectratio]{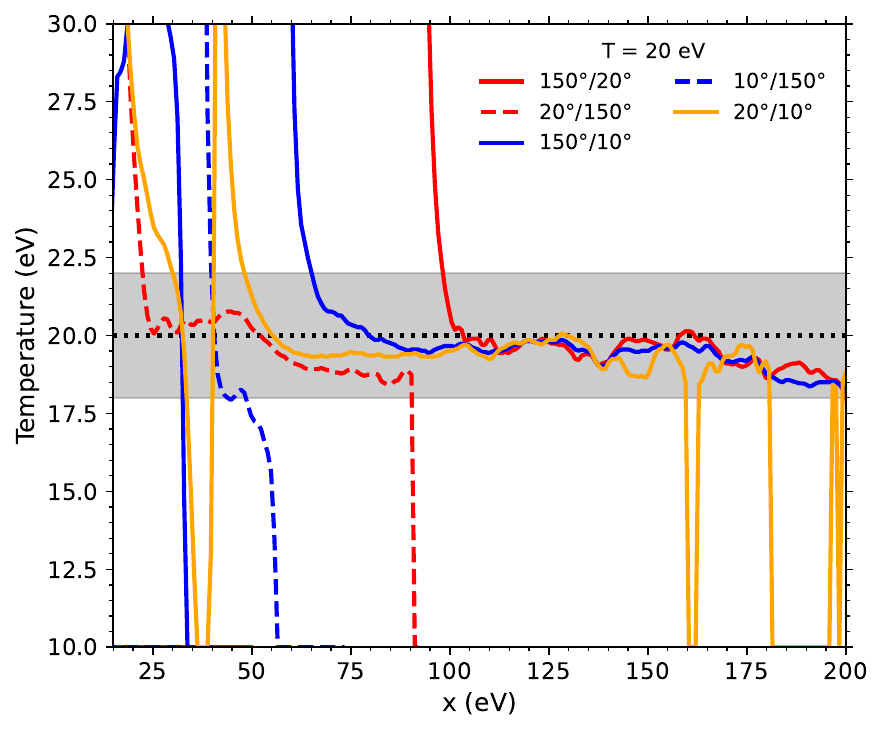}
    \end{subfigure}
    \vspace{0.1mm}
    \begin{subfigure}
        \centering
        \includegraphics[width=0.47\linewidth,keepaspectratio]{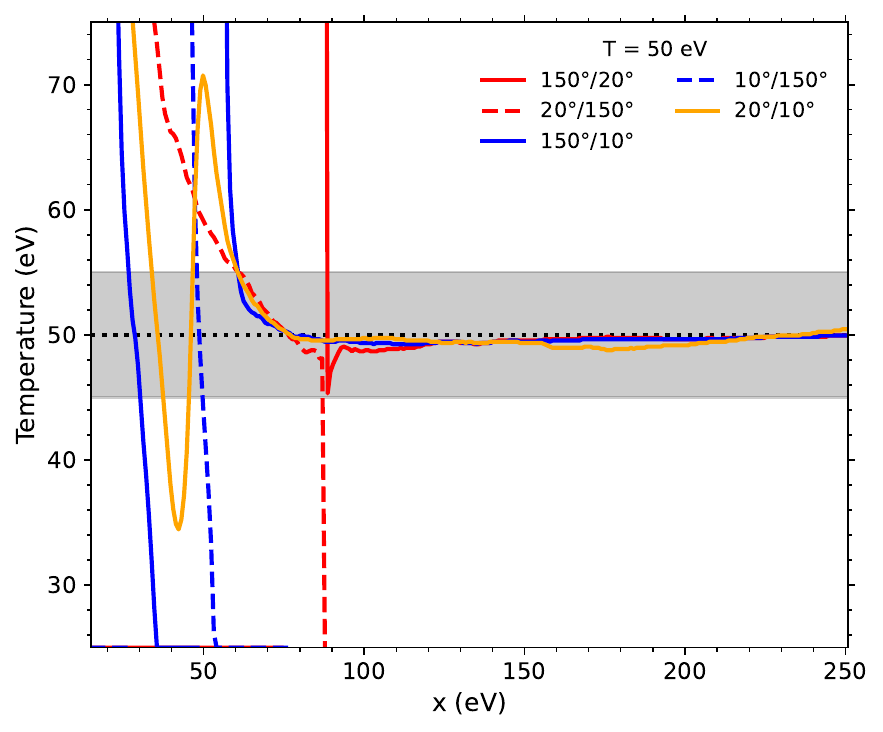}
    \end{subfigure}
    ~
    \begin{subfigure}
        \centering
        \includegraphics[width=0.47\linewidth,keepaspectratio]{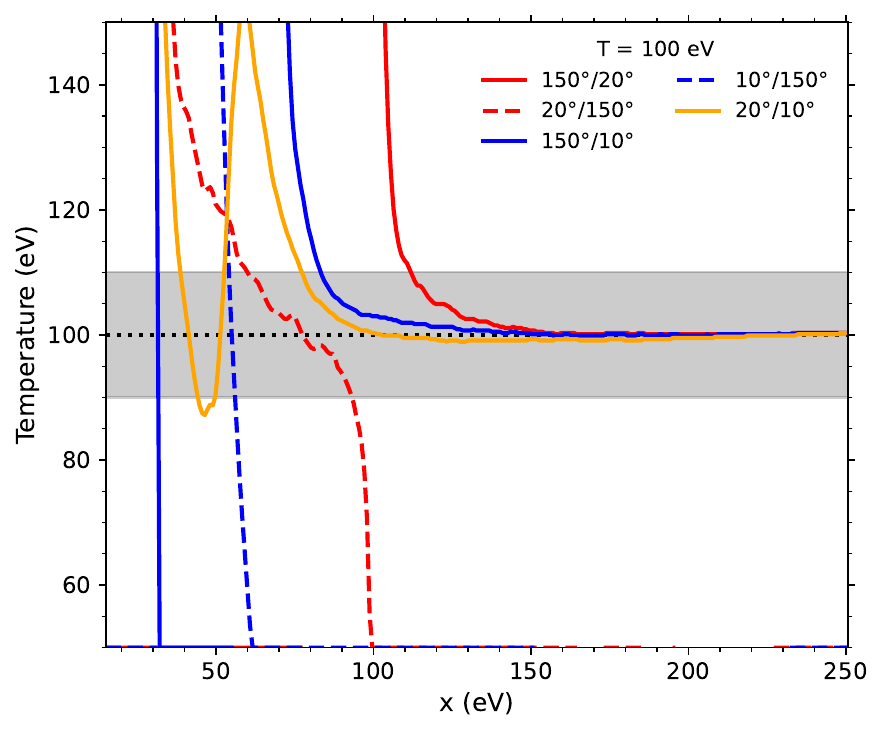}
    \end{subfigure}
    
    \caption{Convergences of the inferred temperature from the minimum of the ITCF ratios versus the integration range $x$ for $T=15$, 20, 50, and 100~eV from the various ratios of the Laplace transformed XRTS spectra at 10$^\circ$, 20$^\circ$, and 150$^\circ$. Solid lines show the higher angle in the numerator. Dashed lines show the lower angle in the numerator, which is equivalent to using the maximum point when the higher angle is in the numerator.
    For clarity, we only plot curves which pass within the temperature range limits of the plots.
    The shaded grey area represents the $T\pm10$\% region around the exact temperature (black dotted).
    }
    \label{fig:NoSIFDeconv}
\end{figure*}

\begin{figure}
    \centering
    \includegraphics[width=0.47\textwidth,keepaspectratio]{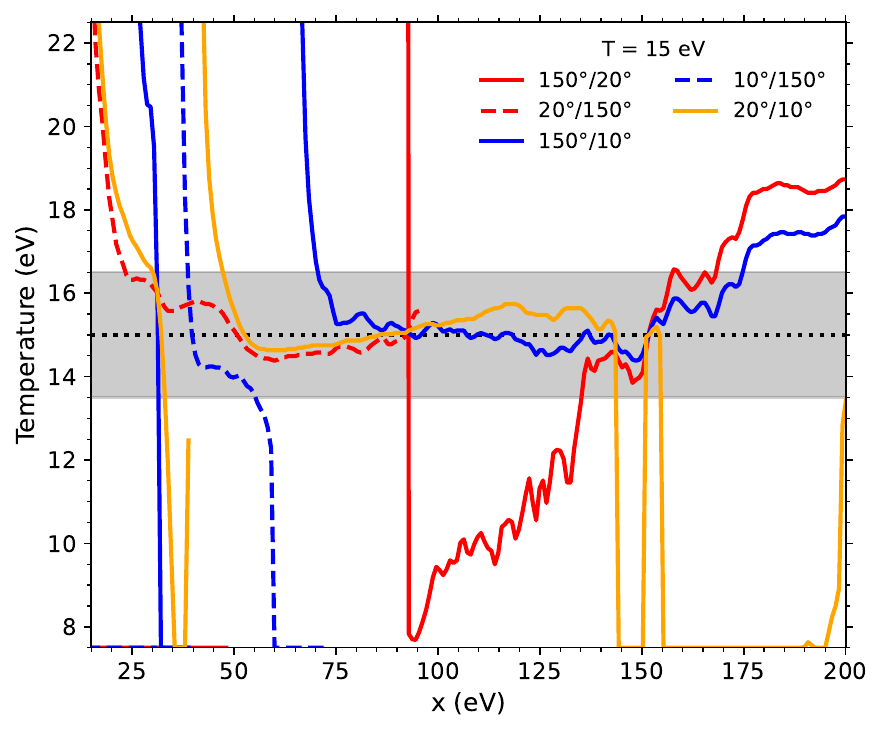}
    \caption{Convergences of the inferred temperature from the minimum of the ITCF ratios versus the integration range $x$ for $T=15$~eV from the various ratios of the Laplace transformed XRTS spectra at 10$^\circ$, 20$^\circ$, and 150$^\circ$. Solid lines show the higher angle in the numerator. Dashed lines show the lower angle in the numerator, which is equivalent to using the maximum point when the higher angle is in the numerator.
    Compared to Fig.~\ref{fig:NoSIFDeconv}, the Rayleigh weight has been reduced by a factor of 4 here.
    The shaded grey area represents the $T\pm10$\% region around the exact temperature (black dotted).
    }
    \label{fig:NoSIFDeconv_RedWr}
\end{figure}

Confident in the treatment of the SIF as approximately a convolution within the spectral range considered here ($x \le 250$~eV), we may now investigate the ITCF ratio method to extract temperature. To start, we presume the SIF to be the same between all scattering angles, though differences between the crystals will be investigated later. Also note that throughout, the integration interval $x$ is the same for both the numerator and denominator. Fig.~\ref{fig:NoSIFDeconv} shows the temperatures inferred from the various ratios of the Laplace transformed spectra for scattering angles of 10$^\circ$, 20$^\circ$, and 150$^\circ$ for each temperature. For clarity, we only examine the minima of the ratios since inverting the ratio is the equivalent of inferring the temperature using the maximum point. We also only plot ratios that have inferred temperatures in the temperature range of the plot limits ($T_{\rm inf} \in [0.5\times T_{\rm exact}, 1.5 \times T_{\rm exact}]$).

For the $T=20$~eV, 50~eV, and 100~eV cases, there are three ratios which show convergence, all with the higher scattering angle in the numerator of the ITCF ratio.
The $T=50$~eV and 100~eV cases in particular produce very accurate temperatures over a very large range of $x$ due to their large upshifted signal.
We also note these two high temperatures show more accurate temperature inferences using the ratio method in Fig.~\ref{fig:NoSIFDeconv} than using the standard ITCF method in Fig.~\ref{fig:DirectDeconv}.
The other ratios do not show meaningful, if any, convergence -- however, these can simply be discarded since their inverses do show convergence (or, equivalently, one could look at their maximum points). For $T=20$~eV, at $x>160$~eV, there are drops in the inferred temperature to $T_{\rm exact}/2$ which, as with ITCFs for Fig.~\ref{fig:DirectDeconv}, indicates the ITCF ratio has lost its well-defined minimum. 

For $T=15$~eV, only one ITCF ratio, $20^\circ/10^\circ$, shows a region of convergence in its minimum that could be used to extract temperature; the remainder do not yield useful information. This is because of the very low number of upshifted photons measured. Indeed, the situation is worse than in Fig.~\ref{fig:DirectDeconv} since we now require upshifted signal in two ITCFs in order to measure a temperature. If the Rayleigh weight is reduced by a factor of 4, the ratio method is more successful at extracting the temperature, as shown in Fig.~\ref{fig:NoSIFDeconv_RedWr}. In this case, there are three ratios -- $20^\circ/150^\circ$, $150^\circ/10^\circ$, and $20^\circ/10^\circ$ -- which have a region of convergence in their minima that can infer the temperature. When $x > 140$~eV, these minima of these ratios begin to diverge or collapse to $T_{\rm exact}/2$. The 15~eV results again highlight that the ability to measure low temperatures is strongly determined by the number of measured upshifted photons, which from detailed balance becomes exponentially more difficult as the temperature decreases. For the HAPG crystal used here, the problem is made worse by the extended crystal IF, meaning upshifted photons need to be measured relatively far from the elastic peak in order to observe convergence. Low-mosaicity HOPG crystals may therefore be more useful for low temperature measurements since the Gaussian distribution of crystallites reduces the extent of the crystal broadening~\cite{Gerlach_JAC_2015}.

Figs.~\ref{fig:NoSIFDeconv} and~\ref{fig:NoSIFDeconv_RedWr} also highlight the strength of collecting XRTS spectra at three or more scattering angles: the analysis of the convergence is made much simpler by being able to cross-check the extracted temperatures between three or more ratios.
At the same time, for the $T=20$--100~eV cases, the high consistency in the inferred temperatures suggests that discrepancies of even a few eV in the inferred temperatures from the ratios of three or more scattering angles would be a strong indication of non-equilibrium effects in the system~\cite{Vorberger_Noneq,Bellenbaum_2025_Warm}.
However, as indicated by the $T=15$~eV results, the ability to do so is closely tied to the number of measured photons, and this caveat is explored in more detail in the following Section~\ref{sec:PhotonCounts}.

\subsection{Robustness to Spectral Noise}\label{sec:PhotonCounts}

\begin{figure*}
    \centering
    \includegraphics[width=\linewidth,keepaspectratio]{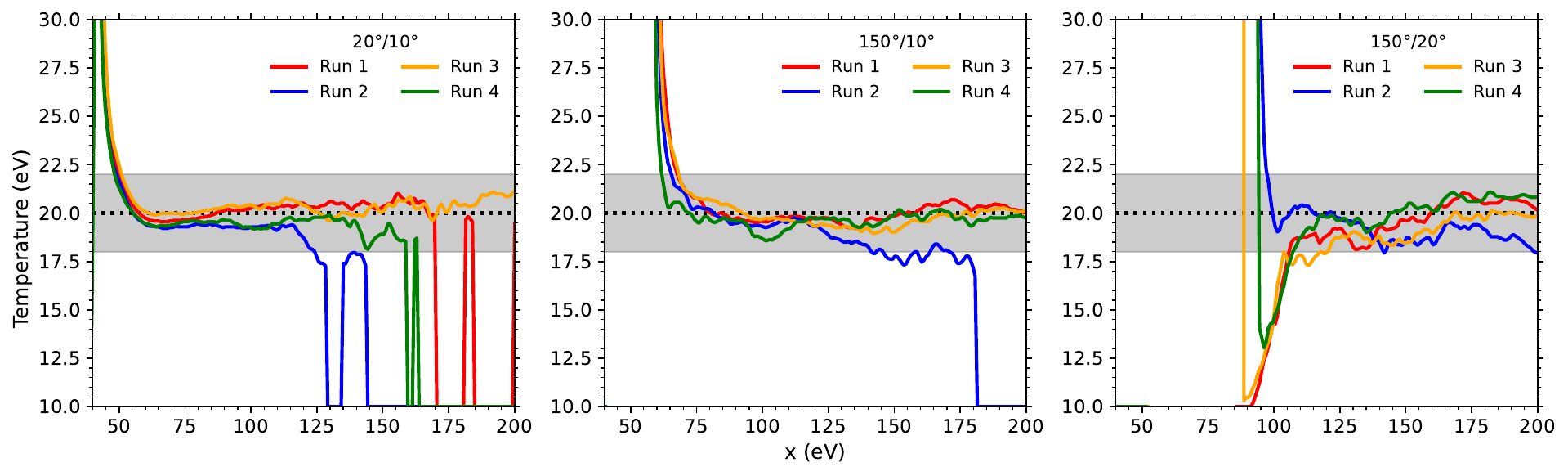}
    \caption{Temperature convergences for $T = 20$~eV for three different scattering angle ratios -- 20$^\circ$/10$^\circ$ (left), 150$^\circ$/10$^\circ$ (centre), 150$^\circ$/20$^\circ$ (right) -- from different ray tracing simulations when $\sim2.5 \times 10^6$ photons have been measured within $x=200$~eV.}
    \label{fig:SpecNoise}
\end{figure*}

\begin{figure*}
    \centering
    \includegraphics[width=\linewidth,keepaspectratio]{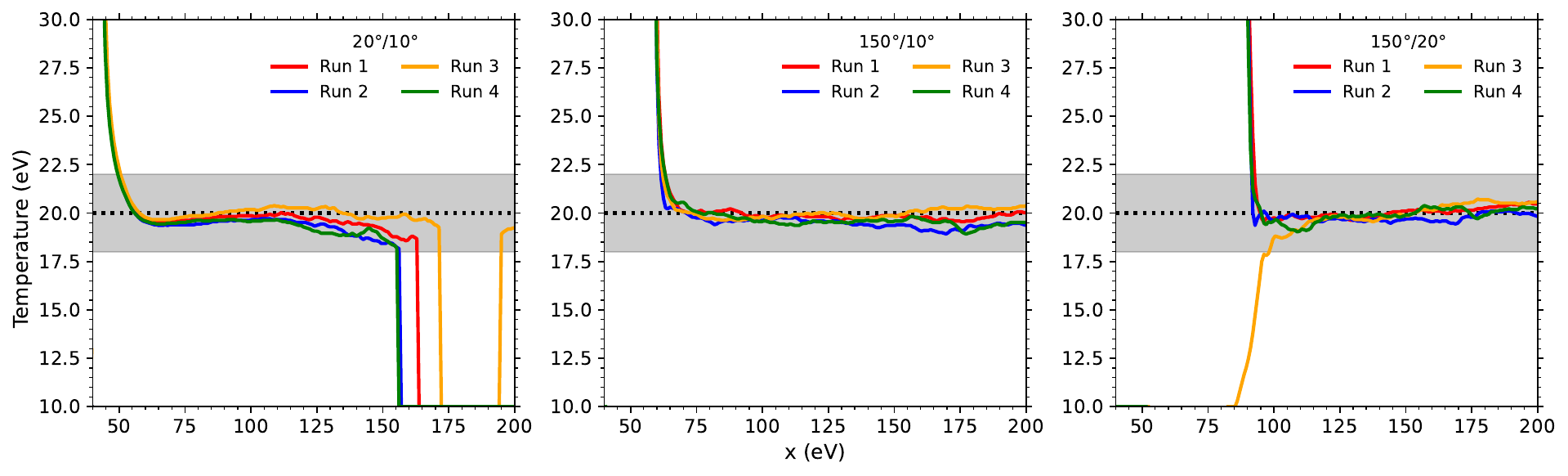}
    \caption{The same as Fig.~\ref{fig:SpecNoise}, but with the Rayleigh weight reduced by a factor four. The various convergences are more similar to each other due to the increase in measured upshifted photons.
    }
    \label{fig:SpecNoiseRedWr}
\end{figure*}

\begin{figure*}
    \centering
    \begin{subfigure}
       \centering
       \includegraphics[width=\linewidth,keepaspectratio]{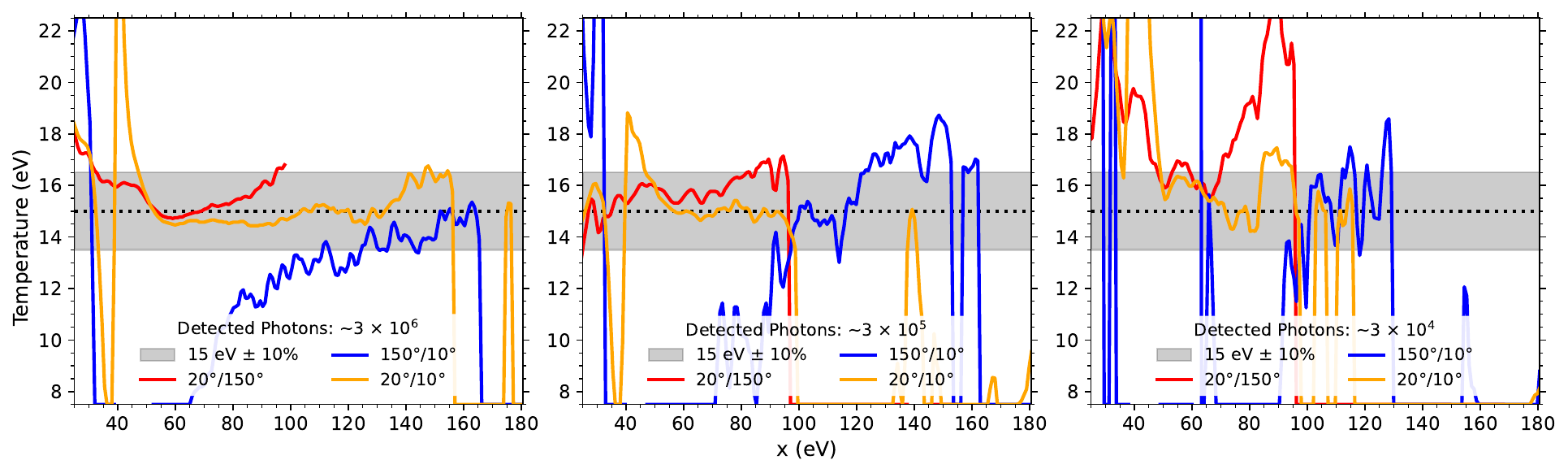}
    \end{subfigure}
    \vspace{0.1mm}
    \begin{subfigure}
        \centering
        \includegraphics[width=\linewidth,keepaspectratio]{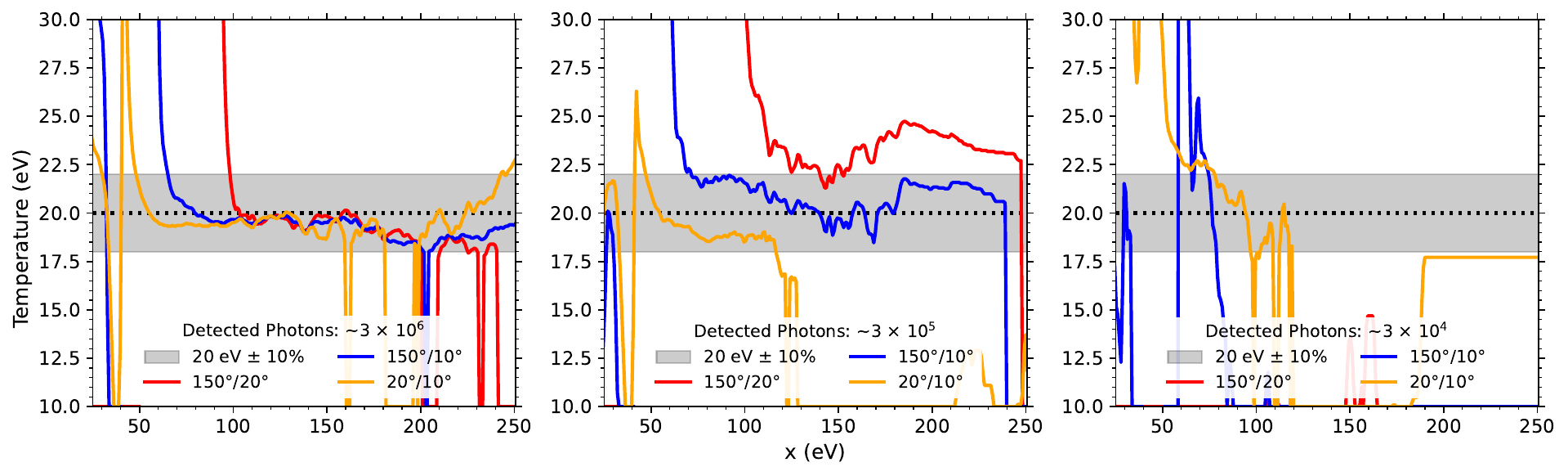}
    \end{subfigure}
    \vspace{0.1mm}
    \begin{subfigure}
        \centering
        \includegraphics[width=\linewidth,keepaspectratio]{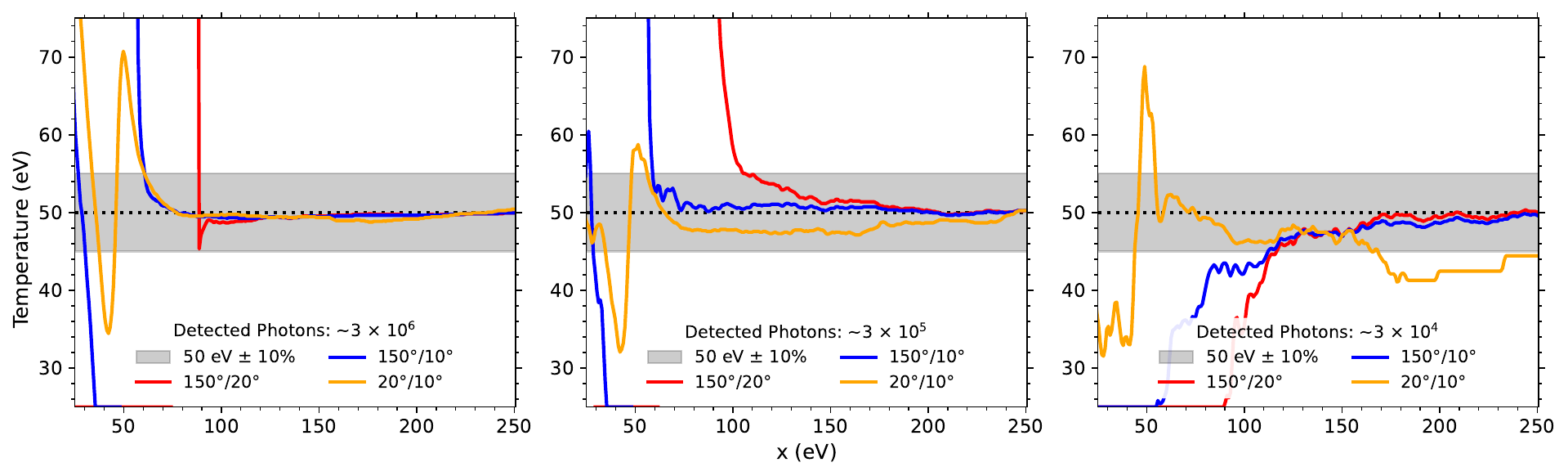}
    \end{subfigure}
    \vspace{0.1mm}
    \begin{subfigure}
        \centering
        \includegraphics[width=\linewidth,keepaspectratio]{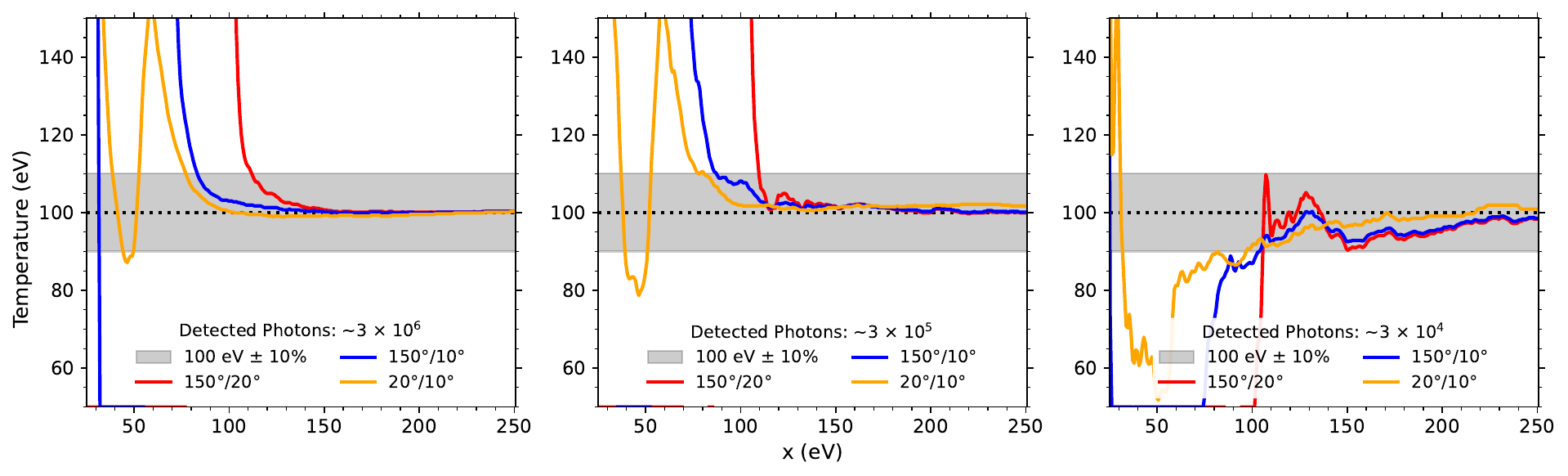}
    \end{subfigure}
    
    \caption{Temperature convergences for $T = 15$~eV, 20~eV, 50~eV, and 100~eV (from top to bottom) for different numbers of photons measured in the spectrum -- the original counts of $3\times 10^6$ (left), $3\times 10^5$ (centre), and $3\times 10^4$ (right) -- for different scattering angle ratios (indicated by the colours).}
    \label{fig:PhotonStats}
\end{figure*}

A reality of experiments is that there will be spectral noise, which introduces uncertainty into inferred results, both for the ITCF method~\cite{Dornheim_T_2022,Dornheim_T2_2022} and for forward modelling~\cite{Kasim_PoP_2019,Boehme_2023_FreeBound,Tilo_Nature_2023}. It is therefore pertinent to check whether the convergence behaviour of a given set of ratios depends strongly on the spectral noise of the experiment.

First, it is important to check that the results do not strongly depend on the specific spectrum sampled at a fixed photon count. Since HEART is a Monte Carlo ray tracer, different results are produced by changing the seed for the random number generator. Each of the ray tracing simulations shown so far have been run with different seeds, so the fact that temperature convergences are already observed is a positive indication of the noise stability. To test the influence of noise further, we compare ratios between sets of spectra in Fig.~\ref{fig:SpecNoise}, where each spectrum is again generated with a fresh seed. At these photon statistics, very similar convergence behaviour is observed, although there are some differences. Reducing the Rayleigh weight (see Fig.~\ref{fig:SpecNoiseRedWr}) or collecting more photons results in the convergences being even more similar to each other.

The results will also depend on the number of photons collected in a given spectrum. The spectra shown in Fig.~\ref{fig:Spectra} contain $\sim3 \times 10^6$ photons in total.
At XFEL facilities, the high repetition rate of the XFEL (up to 10~Hz per frame at the European XFEL, containing potentially multiple bunches at 4.5~MHz~\cite{Zastrau_2021}; and up to 120~Hz per frame at the LCLS~\cite{glenzer2016matter}), means that it is possible to collect tens of thousands of shots very quickly. The high degree of stability of XFEL pulses, combined with recent developments in stable high repetition rate (1-10~Hz) laser drive systems such as DiPOLE-100X~\cite{DiPOLE100} and ReLaX~\cite{ReLaX}, means that collecting the sorts of photon statistics shown here is achievable.
In the case of backlighter sources, such as those used at the National Ignition Facility~\cite{Tilo_Nature_2023,Doeppner_2014_MACS} and OMEGA laser facility~\cite{Fletcher_2014_Observations,Kritcher_2011_In-flight}, the laser heating of the foil results in a tremendous amount of photon emission, such that good photon statistics can be accumulated in single shots (a necessity, given a delay of several hours between shots).

As the number of detected photons can depend on a variety of factors (e.g. beam performance, transmission through the beamline optics, experiment time, the chosen spectrometer, etc.), we have also examined the effect of reducing the number of photons measured on the inferred temperatures.
Examples are shown in Fig.~\ref{fig:PhotonStats} for the four different temperatures, where the input photons have been reduced by a factor of 10 and 100. For the figures shown here, we only plot ratios whose minima showed convergence behaviour, or the potential to do so.

For $T=15$~eV, when the number of photons is reduced by a factor of 10, it is still possible to infer temperatures from the minima of the $20^\circ/10^\circ$ ratio from its convergence between $x=60$--90~eV, while the other two ratios show increased noise. The results likely have some dependence on the specific spectrum sampled due to the reduced photon counting statistics. Reducing the photon counts by another factor of 10 means it is no longer possible to infer a temperature, a consequence of too few upshifted photons being measured. In all three cases, inferring non-equilibrium is not possible unless alternative scattering angles are considered or more photons are collected.

For $T=20$~eV, when the number of photons is reduced by a factor of 10, temperatures may still be inferred from the different ratios, however there are now differences in the three inferred temperatures, and the uncertainty on these temperatures is increased to $\sim2$~eV. Within the uncertainty indicated by the spread of the convergence regions, there is overlap between the temperatures. However, the results also likely have some dependence on the specific spectrum sampled due to the increased photon counting noise. It would also be challenging to infer any non-equilibrium effects unless there are temperature discrepancies of at least 10~eV.
Reducing the photon counts by another factor of 10 reduces the upshifted signal sufficiently that detailed balance cannot be probed, and a temperature cannot be extracted.
There is a completely flat region at $x>190$~eV for the $20^\circ / 10^\circ$ ratio, however this convergence is spurious and is because of the low photon detections at large energy shifts. Additionally, an accurate, converged inferred temperature would not be expected to suddenly emerge after the collapse of the ratio seen between $x = 120$--190~eV. Therefore this apparent temperature can be ignored.

For the $T=50$~eV case, the increased number of photons in the upshifted side of the spectrum due to the high temperature means that temperatures can be inferred using the three ratios, even with a 100 fold reduction in the number of detected photons.
When the photons have been reduced by a factor of 100, there appear to be two different regions of convergence for the $20^\circ/10^\circ$ ratio: between $x \in [90, 150]$~eV and $x \in [200, 230]$~eV.
In the latter case, there is again a completely flat behaviour, and this unusual behaviour could be used to justify discarding the temperature inferred in this region. Discarding this region would also be supported by the consistency of the first region's temperature with the other two scattering angles (again, highlighting the benefit of a three scattering angle setup). Alternatively, one could interpret the two convergence regions as providing a large uncertainty estimate on the inferred temperature at this ratio. Regardless, the large upshifted signal indicates that even lower photon numbers could be acceptably used to infer temperatures.

Finally, at $T=100$~eV, the upshifted signal is even stronger. Consistent temperatures are observed for all three ratios at the three different signal levels, although the difference and uncertainty increases up to $\sim5$~eV for the lowest photon counts considered. In this case, non-equilibrium effects could be detected if temperature discrepancies of several eV are observed.

Once again, the results shown in this section strongly depend on the Rayleigh weight: reducing the Rayleigh weight to increase the portion of photons scattered inelastically means that fewer total photons can suffice for a given temperature, and the temperature discrepancies required to infer non-equilibrium are likewise lowered. Since the ratio of inelastic to elastic scattering depends on the specific system being probed, planning is required to determine the number of photons that must be detected and the optimal choice of scattering angles in order to adequately infer the temperature or non-equilibrium from the ITCF ratios.

\subsection{Differences in Spectrometer Alignment}\label{sec:Alignment}

\begin{figure*}
    \centering
    \begin{subfigure}
        \centering
        \includegraphics[width=\linewidth,keepaspectratio]{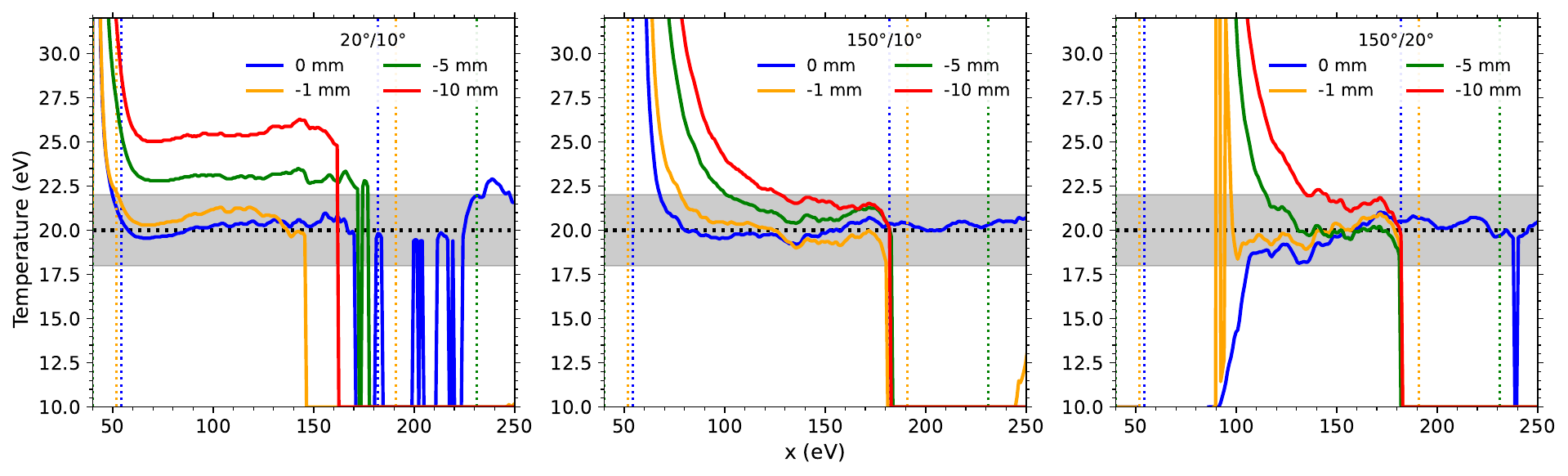}
    \end{subfigure}
    \vspace{0.1mm}
    \begin{subfigure}
        \centering
        \includegraphics[width=\linewidth,keepaspectratio]{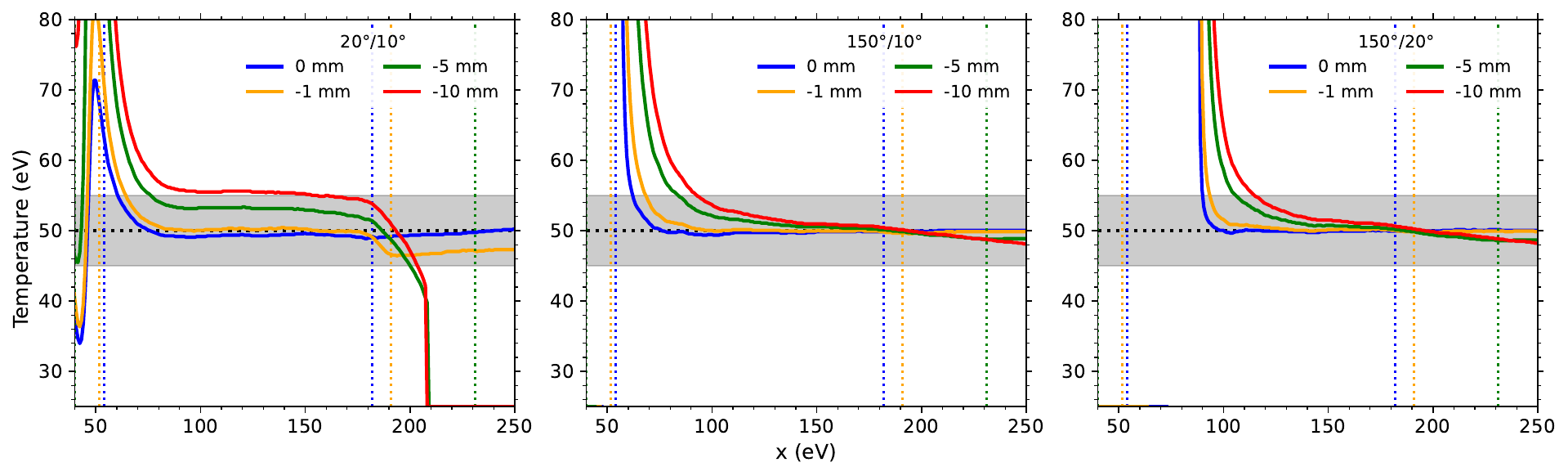}
    \end{subfigure}
    \caption{Temperature convergences for $T = 20$~eV (top) and 50~eV (bottom) for three different scattering angle ratios, with one of the spectrometers moving along the dispersion axis by $\Delta l = 0$~mm (blue), -1~mm (orange), -5~mm (green), and -10~mm (red). The positions of the physical crystal edges as they appear in the SIFs of the different spectrometer positions (see Fig.~\ref{fig:ExampleIFs}~(b)) are indicated by the vertical dotted lines of the same colour.
    }
    \label{fig:SpecShift}
\end{figure*}

\begin{figure*}
    \centering
    \includegraphics[width=\linewidth,keepaspectratio]{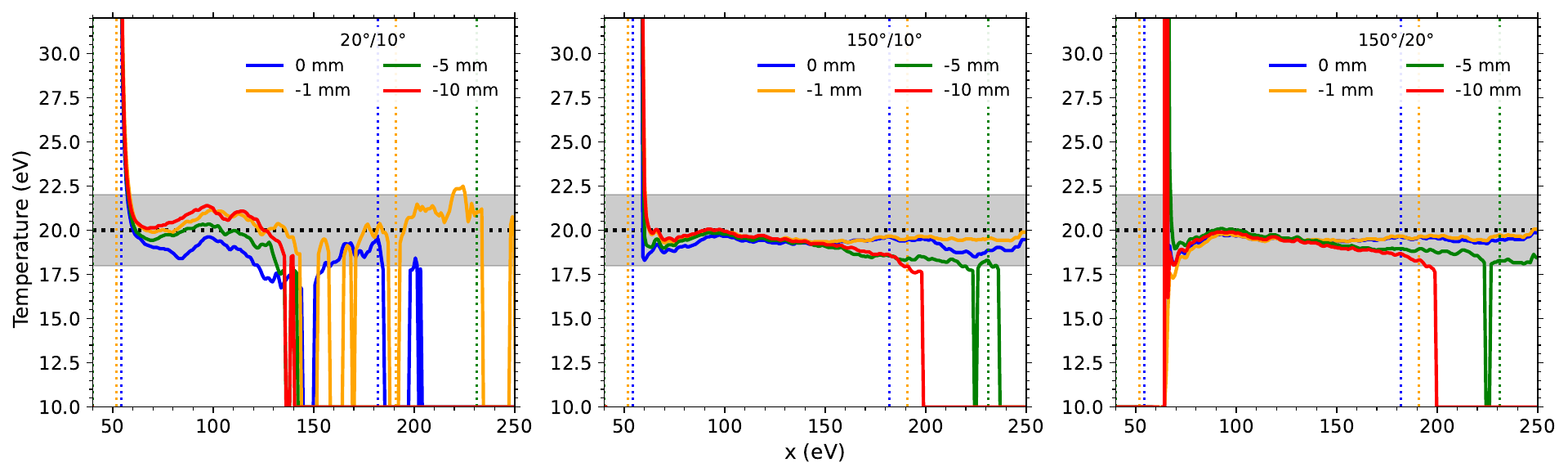}
    \caption{The same as Fig.~\ref{fig:SpecShift} for $T=20$~eV, except the Rayleigh weight is now zero (i.e. no elastic scattering), but the number of inelastically scattered photons is the same as in Fig.~\ref{fig:SpecShift}.
    }
    \label{fig:SpecShift0Wr}
\end{figure*}

\begin{figure*}
    \centering
    \includegraphics[width=\linewidth,keepaspectratio]{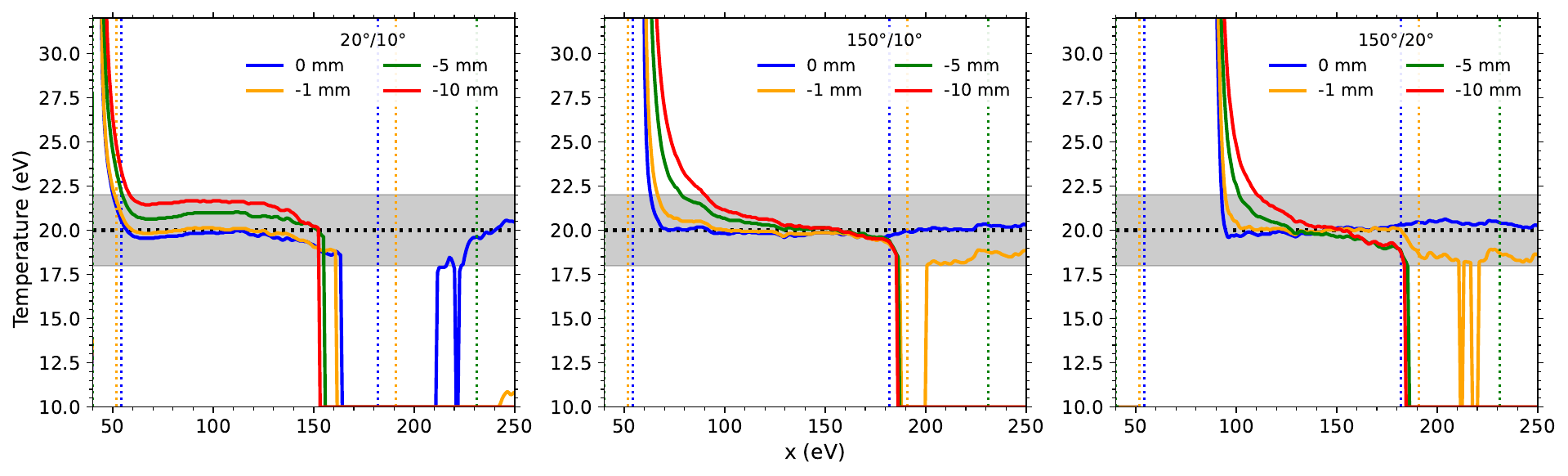}
    \caption{The same as Fig.~\ref{fig:SpecShift} for $T=20$~eV, except the Rayleigh weight has been reduced by a factor of 4.
    }
    \label{fig:SpecShiftRedWr}
\end{figure*}

A common scenario in experiment is that two spectrometers will be at slightly different distances from the target, for example due to spatial limitations in the chamber, or ensuring that key features do not lie on a detector gap. Here we examine whether the results will be strongly affected by such misalignments. The main expected difference is that since the position of the crystal ends on the detector will be in different places for the spectrometers (see Fig.~\ref{fig:ExampleIFs}~(b)), the two SIFs will suddenly no longer resemble each other when an edge is crossed, leading to divergences in the temperature convergence.

Since we are considering von H\'amos spectrometers, which have a clear focal point that is optimised in experiment, it is reasonable to assume that the crystal position below the source-detector axis is identical between spectrometers. We therefore focus on shifts of the crystal position along the dispersion axis, $\Delta l$. For these ratios, one of the crystals is fixed in the original position ($\Delta l= 0$~mm), while the other is moved by $\Delta l= -1$, -5, and -10~mm (i.e. closer to the source; see Fig.~\ref{fig:SpecSetup}). The detector is kept fixed in position since this makes comparisons of the ratios simpler by maintaining the spectral range of the spectrometer.

The temperature convergence for $T=20$~eV and 50~eV for three ratios of scattering angles and different crystal shifts is shown in Fig.~\ref{fig:SpecShift}. In all cases, the shifting spectrometer is in the denominator of the ratio because the ratios' minima did not converge when the shifting spectrometer was in the numerator.
Convergences can be seen in all cases, however there are temperature discrepancies that generally grow with $\Delta l$. In particular, the $20^\circ/10^\circ$ ratio for both $T=20$~eV and $T=50$~eV -- where the where the Rayleigh weights are largest -- show the largest temperature discrepancies. The $T=50$~eV discrepancies are relatively smaller than for $T=20$~eV.

The reason of the temperature discrepancies can be largely ascribed to the elastic feature: Fig.~\ref{fig:SpecShift0Wr} shows the convergences for $T = 20$~eV when the Rayleigh weight has been set to zero (i.e. no elastic scattering), but the number of inelastically scattered photons has been kept the same as the $T=20$~eV case in Fig.~\ref{fig:SpecShift}. The convergence behaviours for the different ratios are now much more consistent with each other, with the spectrometer shift having very little effect, especially for the $150^\circ/10^\circ$ and $150^\circ/20^\circ$ ratios. Likewise, Fig.~\ref{fig:SpecShiftRedWr} shows convergence plots for $T=20$~eV using the same number of input photons as Fig.~\ref{fig:SpecShift}, but with the Rayleigh weight reduced by a factor of four: again, the convergence behaviours are much more similar to one another, but not as similar as the case of no Rayleigh weight. In comparison to the zero Rayleigh weight case, the increased number of inelastically scattered photons in Fig.~\ref{fig:SpecShiftRedWr} means the convergence plots are smoother than in Fig.~\ref{fig:SpecShift0Wr}, but the presence of an elastic feature still leads to larger discrepancies, particularly in the $20^\circ/10^\circ$ when the Rayleigh weights are largest.

The elastic scattering has such a dramatic effect because it also needs to be deconvolved from the spectrum, even though it contains no information about the detailed balance. But, since it is a very prominent feature with sharp edges, discrepancies in the SIFs between the ITCF ratios result in discrepancies in the inferred temperature due to the elastic feature being erroneously handled. The larger the elastic feature is relative to the inelastic scattering, the more its error contributes to the inferred temperature. The likely reason the SIF discrepancies are less problematic for deconvolving the inelastic spectrum alone, as evidenced by Fig.~\ref{fig:SpecShift0Wr}, is that the SIF edges are essentially smoothed over due to their positions changing with energy (see Fig.~\ref{fig:ExampleIFs}~(a)), so the underlying inelastic SIFs look more similar to one another in the ratios.

The effect of the crystal edges in the elastic scattering on the inferred temperature convergence can also be observed by sudden changes in the convergence behaviour, which is easiest to see for the $20^\circ/10^\circ$ ratio at $T=50$~eV in Fig.~\ref{fig:SpecShift}.
The first edge for all crystals occur between $\sim27$--54~eV, but the onset of convergences happens at $x>60$~eV -- at this point, the wings of the SIFs resemble each other fairly closely, which allows for a temperature convergence to be observed, albeit at different temperatures. The second set of edges occur at $x>180$~eV, starting with the $\Delta l=0$~mm edge (indicated by the blue vertical dotted line). Once this edge is crossed, the inferred temperatures begin to drop due to two 8.5~keV SIFs in the ratio no longer resembling each other (see Fig.~\ref{fig:ExampleIFs}~(b)). However, once the second crystal's edge is crossed (e.g. for the $\Delta l = -1$~mm case), it is possible for the convergence to stabilise again, provided it has not been lost entirely (e.g for the $\Delta l=-5$~mm and $\Delta l=-10$~mm cases). For the $150^\circ/10^\circ$ and $150^\circ/20^\circ$ cases at $T=50$~eV, convergence can be re-established in the $\Delta l=-1$~mm and $\Delta l=-5$~mm cases. Since the second edge for the $\Delta l=-10$~mm shift does not fit in the spectral range, this convergence does not have a chance to be re-established. Observing the edges of the crystal in the convergence plots may be used then to restrict the range of $x$ used to infer temperature in cases where two different temperatures are implied in the same convergence plot.

Overall, the results in this section suggest that the spectrometers do not necessarily need to be perfectly aligned with one another, however the acceptable misalignment is strongly influenced by the relative strength of the elastic scattering to the inelastic scattering, with stronger inelastic scattering allowing for larger misalignments. Small misalignments of $\Delta l \sim 1$~mm are generally acceptable, and the convergence behaviour is generally similar to perfect alignment. Larger misalignments of up to 5--10~mm can still provide accurate temperature, provided the inelastic signal is sufficiently strong relative to the elastic scattering. Therefore, the tolerance for misalignment depends directly on the system being measured, and this would need to be considered ahead of an experiment. In an experiment, measurements of the IF should be performed to identify at least the first edge of the spectrometers' crystals (which can also be used to locate the position of the second edge~\cite{Gawne_2024_Effects} in case it is too weak to be measured directly) to check whether the crystals are sufficiently aligned for the systems being probed.

\subsection{Differences in the Mosaic Crystal}\label{sec:Mosaic}

We also consider the case that the crystal properties -- namely the mosaicity and rocking curve widths -- are not identical. This is particularly pertinent for mosaic crystals: fundamentally, the distributions describing the crystallites are random, so even nominally similar crystals can show very different instrument functions.

\subsubsection{Mosaicity}

\begin{figure*}
    \centering
    \includegraphics[width=\linewidth,keepaspectratio]{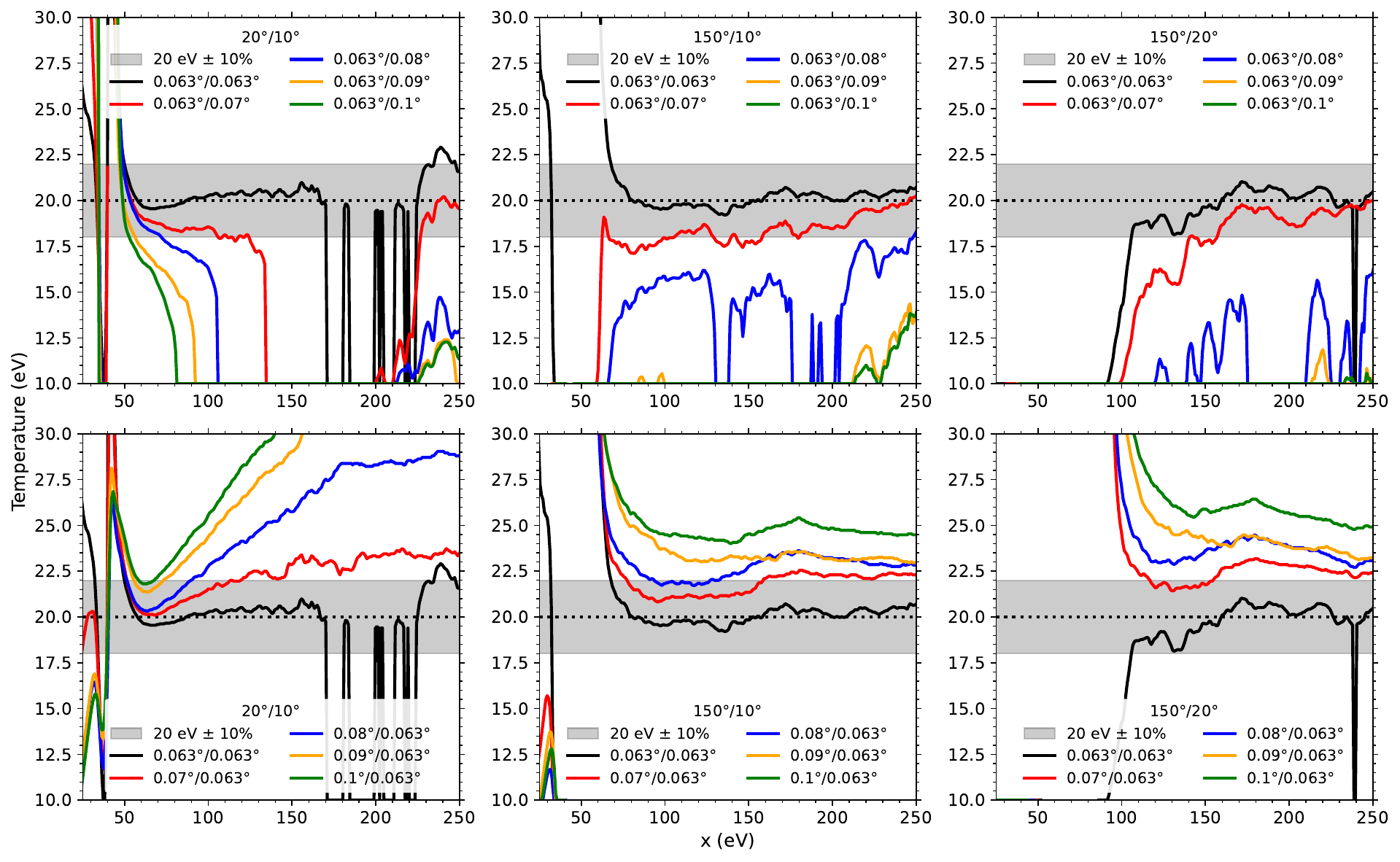}
    \caption{Temperature convergences for $T=20$~eV for three different scattering angle ratios, but with the mosaicity of one of the crystals different from the original 0.063$^\circ$: 0.07$^\circ$ (red), 0.08$^\circ$ (blue), 0.09$^\circ$ (orange), and 0.10$^\circ$ (green). Also shown in black is the original case where the crystals are identical (i.e. 0.063$^\circ$/0.063$^\circ$). The top plots show the cases when the higher mosaicity is in the denominator, and the bottom plots when it is in the numerator. 
    }
    \label{fig:MosaicDifference_20eV}
\end{figure*}

\begin{figure*}
    \centering
    \includegraphics[width=\linewidth,keepaspectratio]{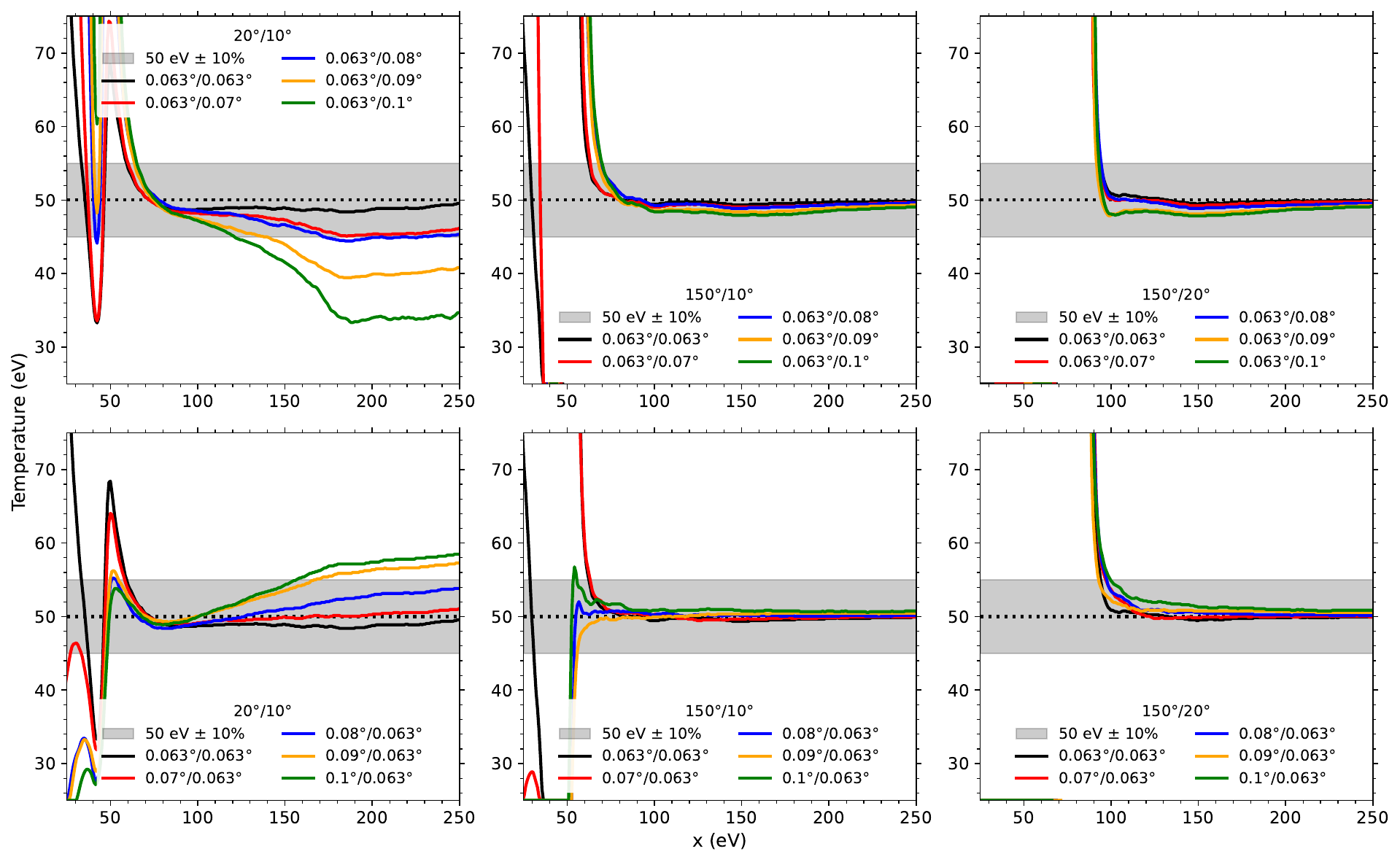}
    \caption{The same as Fig.~\ref{fig:MosaicDifference_20eV}, but for $T=50$~eV.
    }
    \label{fig:MosaicDifference_50eV}
\end{figure*}

\begin{figure*}
    \centering
    \includegraphics[width=\linewidth,keepaspectratio]{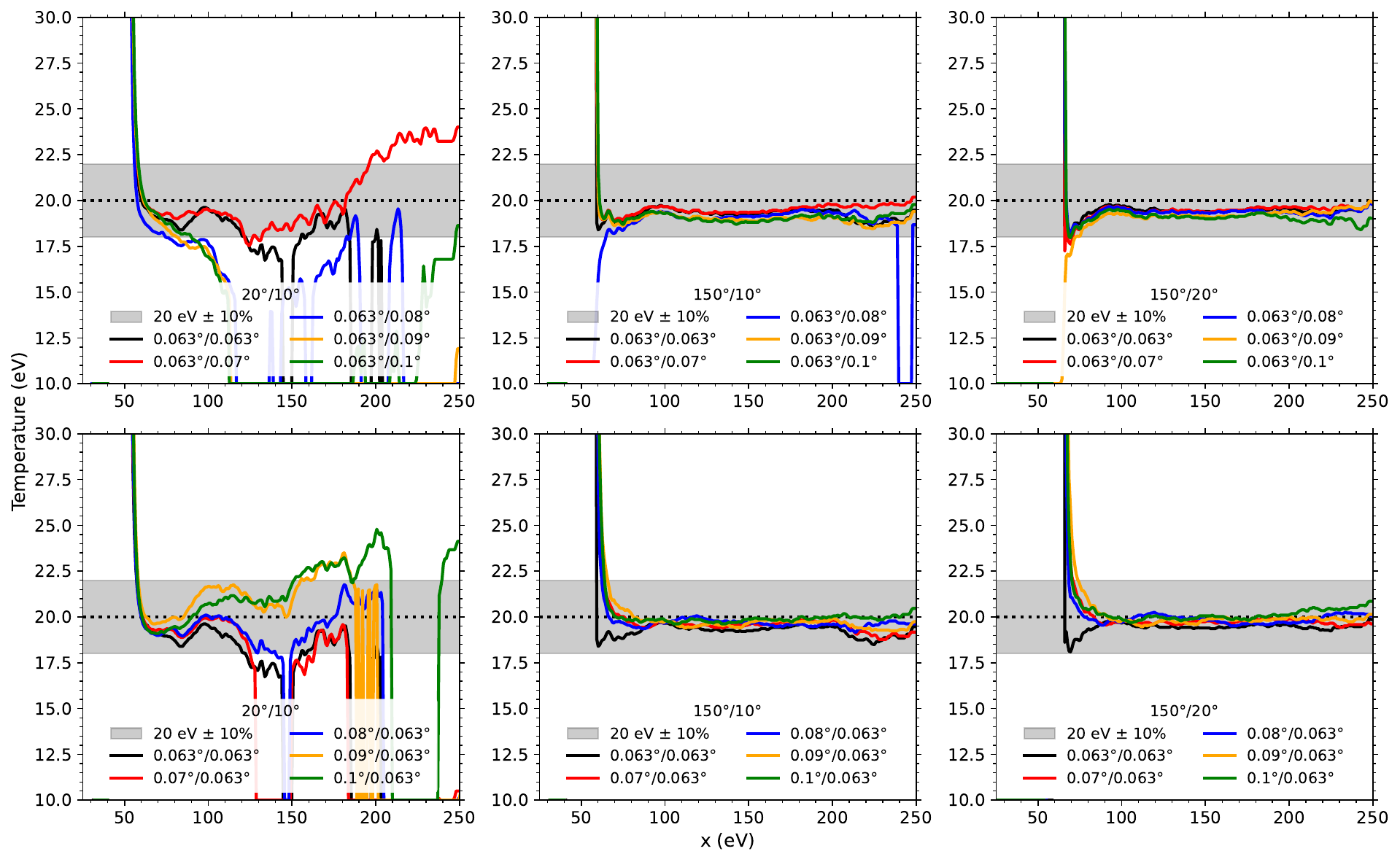}
    \caption{The same as Fig.~\ref{fig:MosaicDifference_20eV} for $T=20$~eV, except the Rayleigh weight is now zero (i.e. no elastic scattering), but the number of inelastically scattered photons is the same as in Fig.~\ref{fig:MosaicDifference_20eV}.
    \label{fig:MosaicDifference0Wr}
    }
\end{figure*}


First we consider the mosaicity of the crystal, which is the width of its mosaic distribution function (MDF), which is the probability distribution function of the crystallite orientation to the surface normal. The main contribution of the MDF to the crystal IF is broadening towards higher photon energies on the detector, both directly~\cite{Gawne_2024_Effects} and by enhancing depth broadening~\cite{Schlesiger_JAC_2017}, resulting in a highly asymmetric IF. The MDF also makes the entire surface of the crystal reflective to all photon energies, which results in the physical edges of the crystal being observed in the IF and in spectra~\cite{Gawne_2024_Effects}; see Figs.~\ref{fig:ExampleIFs} and~\ref{fig:Spectra}. 
Increasing the mosaicity increases the asymmetry, but at the same time it also makes the SIF flatter as the peak reflectivity is reduced~\cite{zachariasen1994theory}. The mosaicity therefore has a substantial effect on the shape of the instrument function and spectra. Given the strong effect of the elastic feature seen in Section~\ref{sec:Alignment}, it is strongly anticipated the elastic scattering will have a noticeable effect on the quality of the convergences observed here.

In Fig.~\ref{fig:MosaicDifference_20eV} and~\ref{fig:MosaicDifference_50eV}, we compare inferred temperature convergence for $T=20$~eV and $T=50$~eV, respectively, where one of the spectrometers has the original 0.063$^\circ$ mosaicity, and the other has increasing mosaicity. Both the case where the denominator spectrum has the increasing mosaicity and the case where the numerator has the increasing mosaicity are plotted. This distinction has a noticeable effect on the convergence behaviour for $T=20$~eV ratios and for the $20^\circ/10^\circ$ ratio at $T=50$~eV -- these are the cases where the elastic scattering is very strong compared to the inelastic scattering. Once again, when the elastic scattering is very prominent, it has a noticeable effect on the convergence due to differences in the 8.5~keV SIF between the two ratios. Removing the elastic feature while keeping the inelastically scattered signal the same results in excellent agreement between the $150^\circ/10^\circ$ and $150^\circ/20^\circ$ ratios regardless of the mosaicity and their position in the ratio, as shown in Fig.~\ref{fig:MosaicDifference0Wr}; a similar observation is made for the $T=50$~eV case in Fig.~\ref{fig:MosaicDifference_50eV}. The $20^\circ / 10^\circ$ case in Fig.~\ref{fig:MosaicDifference_50eV} and~\ref{fig:MosaicDifference0Wr} shows generally poor convergence stability to changes in the mosaicity, indicating that the narrow plasmon peak shapes are strongly influenced by the shape of the SIFs in their vicinity.

The results here demonstrate that differences in mosaicity are acceptable, once again provided that the elastic scattering is sufficiently weak relative to the inelastic scattering, which will be system dependent. If this is the case, even large differences in the mosaicity are acceptable, which is promising since the exact mosaicity of a crystal (or even across a crystal~\cite{Zastrau_JoI_2013}) is fundamentally random. These results also suggest that an ideal setup should use at least one large scattering angle (here $150^\circ$) and a smaller angle (here, $10^\circ$ and $20^\circ$) -- the ratio of the large angle over either of the two small scattering angles provides very similar temperatures to one another for these equilibrium systems, even when the crystals have very different mosaicities. Although the ratio of the two small scattering angles may not yield a converged temperature, comparison of the inferred temperatures from the remaining two ratios using the high scattering angles would be sufficient to either confirm the temperature or identify no equilibrium by differences in temperature of a few eV (again, with the caveat of the relative strength of the elastic feature).

\subsubsection{Intrinsic Rocking Curve Widths}\label{sec:IRC}

\begin{figure*}
    \centering
    \includegraphics[width=\linewidth,keepaspectratio]{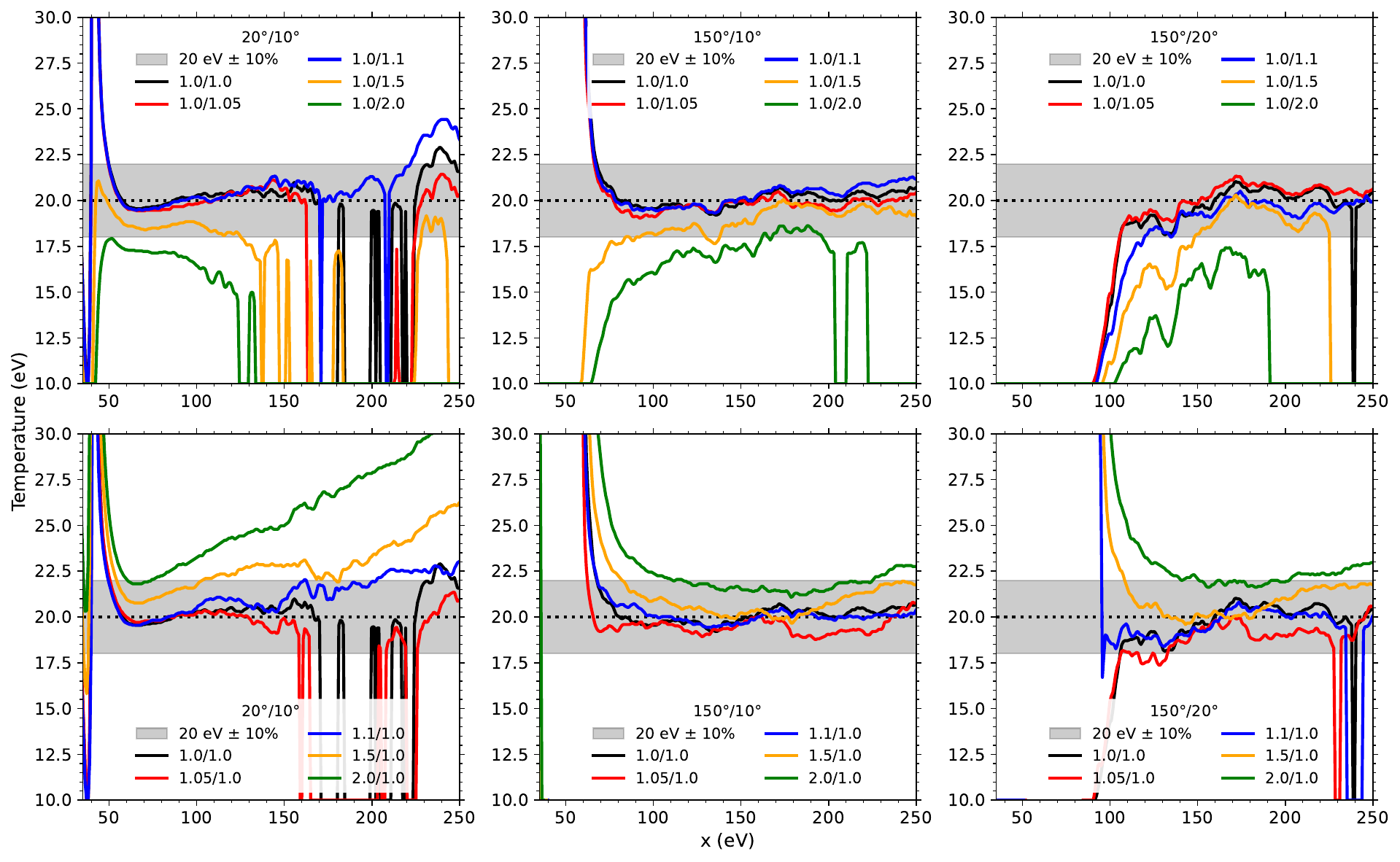}
    \caption{Temperature convergences for $T=20$~eV for three different scattering angle ratios, but with the IRC width of one of the crystals changed from the original by a factor of 1.05 (red), 1.1 (blue), 1.5 (orange), and 2.0 (green). Also shown in black is the original case where both crystals are identical. The top plots show the cases when the wider IRC is in the denominator, and the bottom plots when it is in the numerator. 
    }
    \label{fig:IRC_Differences_20eV}
\end{figure*}

\begin{figure*}
    \centering
    \includegraphics[width=\linewidth,keepaspectratio]{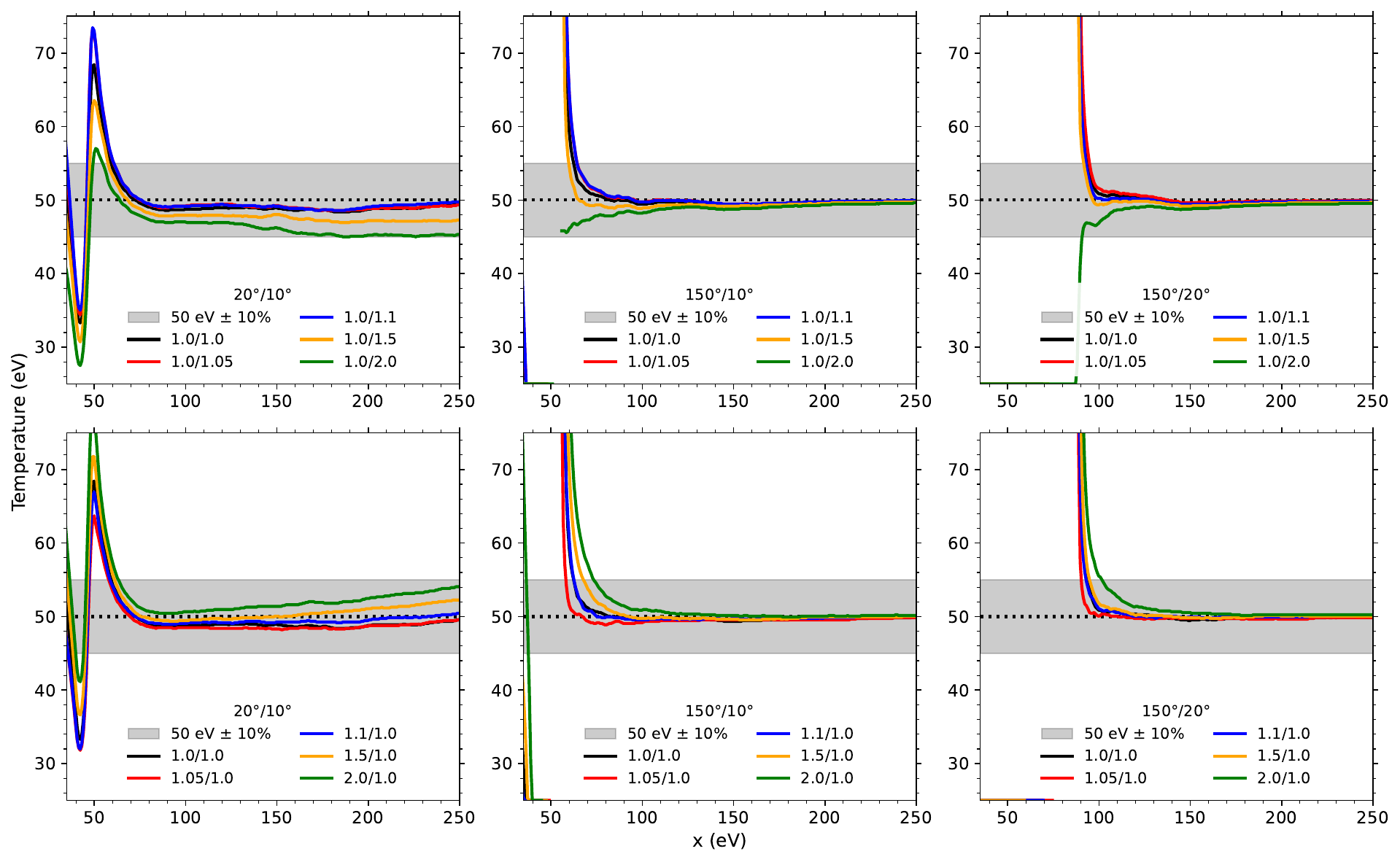}
    \caption{The same as Fig.~\ref{fig:IRC_Differences_20eV}, but for $T=50$~eV.
    }
    \label{fig:IRC_Differences_50eV}
\end{figure*}

\begin{figure*}
    \centering
    \includegraphics[width=\linewidth,keepaspectratio]{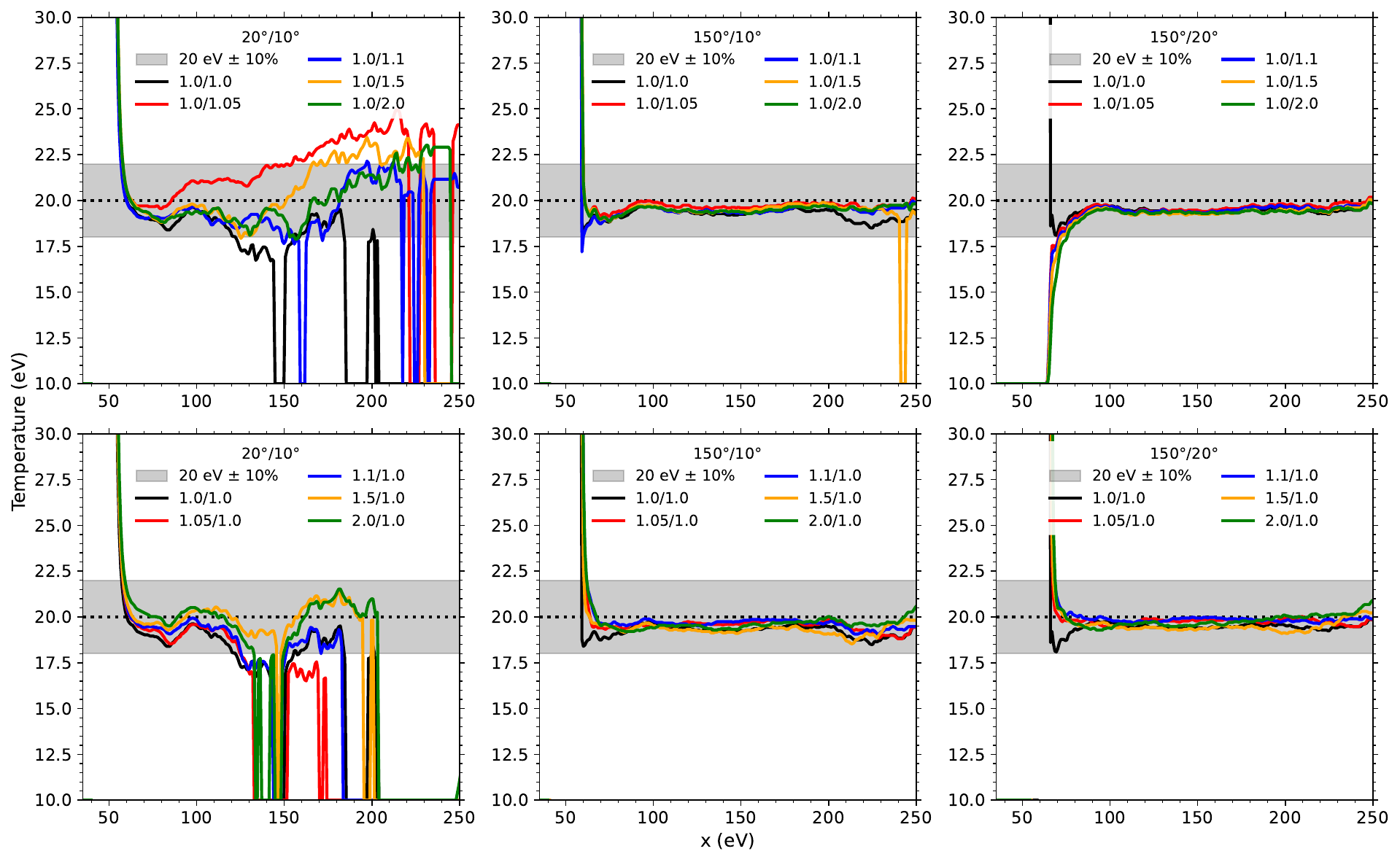}
    \caption{The same as Fig.~\ref{fig:IRC_Differences_20eV}, for $T=20$~eV, except the Rayleigh weight is now zero (i.e. no elastic scattering), but the number of inelastically scattered photons is the same as in Fig.~\ref{fig:IRC_Differences_20eV}.
    }
    \label{fig:IRC_DifferencesNoWr}
\end{figure*}


For a given photon energy interacting with a perfect crystallite, the width of the IRC depends on the thickness of the crystallite~\cite{zachariasen1994theory}.
In testing, we found that the width of the averaged IRC can be very sensitive to the specific crystallite thickness distribution used. Indeed, for the HAPG crystals at the European XFEL there is some indication that the intrinsic rocking curves do have different widths~\cite{Gawne_2024_Effects}.
Therefore, we consider the case where the IRC widths are different between the two crystals.
In isolation, the IRC results in almost-symmetric broadening on the detector, with slight asymmetry towards higher photon energies. Additionally, as with the mosaicity, increasing the width of the IRC results in a flatter reflectivity curve across the crystal due to a reduction in the peak reflectivity.

Within HEART, the best model of the IRC is a Voigt profile, where the user selects the widths of the Gaussian and Lorentzian components. For the energy dependence, the relative Gaussian and Lorentzian character of the Voigt profile is kept the same, but the FWHM of the profile is determined by $W(E)$ in Eq.~(\ref{eq:IRC_FWHM}). For this test, we multiply $W(E)$ by an additional constant factor.

The results are plotted in Fig.~\ref{fig:IRC_Differences_20eV} and~\ref{fig:IRC_Differences_50eV} for $T=20$~eV and 50~eV, respectively. We consider both the case that the broader IRC is in the numerator of the ratio and in the denominator. As with the mosaicity, the situations in which the elastic peak is most prominent (i.e. low temperature and low scattering angles), there is a fairly strong sensitivity to the width of the IRC, however the sensitivity is much lower than for the mosaicity. The $T=50$~eV case shows consistent convergence across all sets of ratios considered. Indeed, even using the standard Rayleigh weight model at $T=20$~eV, the $150^\circ/10^\circ$ and $150^\circ/20^\circ$ ratios give fairly consistent temperatures, especially when the broader IRC crystal is used to measure the low scattering angle spectrum. This points to the fact that planning can be done to decide which crystals of a set should be used to measure the low angle spectrum and the high angle spectrum for optimally using the ratio method to extract temperature.

For completeness, we also show the $T=20$~eV case when there is no elastic scattering but still the same number of inelastically scattered photons in Fig.~\ref{fig:IRC_DifferencesNoWr}. Overall, there is generally an improvement in the consistency of the inferred temperature between the ratios of the crystals, demonstrating once again the importance of the elastic feature to the quality of the convergence. The $20^\circ / 10^\circ$ ratios remain very challenging to use to extract a temperature, but the $150^\circ/10^\circ$ and $150^\circ/20^\circ$ ratios show very high degrees of consistency and accuracy.

Overall, we find that the mosaic crystals do not need to be completely identical for the ratios method to work. This is a promising outcome since the random nature of mosaic crystals means that no two crystals are identical. Moreover, the spectrometer positions do not necessarily need to be identical, which is also promising since in reality it can be difficult to exactly align two spectrometers to the same position. The degree of acceptable difference is highly dependent on the strength of the elastic scattering relative to the inelastic scattering: despite the elastic feature itself containing no useful information about the detailed balance, it still nevertheless needs to be correctly treated otherwise it will lead to errors in the inferred temperature, or prevent a temperature convergence being observed at all. The more prominent the elastic feature is, the more similar the spectrometer instrument functions need to be. Conversely, in systems where the elastic scattering is relatively weak compared to the inelastic scattering, there can be large differences in the spectrometer setup and crystal properties without negatively affecting results. Ultimately, these considerations need to be checked on a system-by-system basis ahead of an experiment.

\subsection{Application to Laser Facility Experiments}\label{sec:NIF}

\begin{figure*}
    \centering
    \includegraphics[width=\linewidth,keepaspectratio]{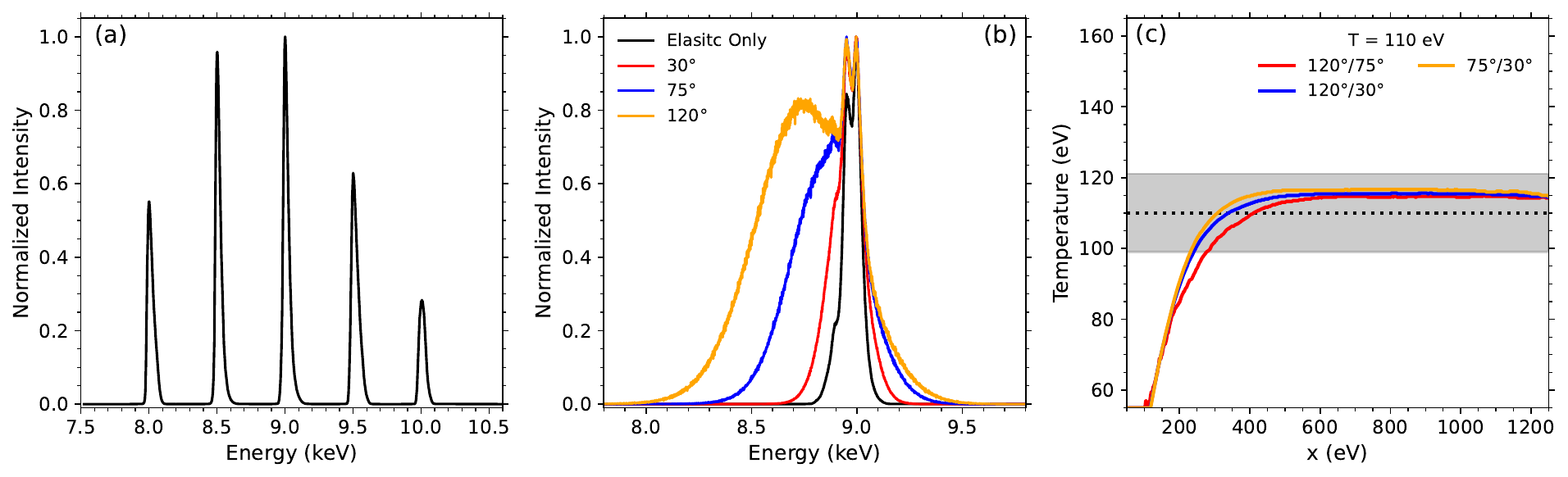}
    \caption{(a) Ray traced instrument functions for 8, 8.5, 9, 9.5, and 10~keV photons on the MACS spectrometer, showing the variation in the instrument function (especially the peak reflectivity) across the spectrometer. (b) Ray traced XRTS spectra of compressed Be at conditions reported in Ref.~\cite{Tilo_Nature_2023} at three different scattering angles. Additionally, the elastic scattering feature is shown to indicate the form of the SIF broadening. (c) Temperature convergences for the three different ratios.
    }
    \label{fig:NIF}
\end{figure*}

So far this work has only considered a narrowband probe beam and relatively high resolution HAPG crystals. However, a number of experimental setups have much larger SIF broadening -- whether that be from lower resolution spectrometers (e.g. the MACS spectrometer at the NIF~\cite{Doeppner_2014_MACS}), or from larger source broadening (e.g. self-amplified spontaneous emission (SASE) beams at XFELs~\cite{kraus_xrts} or backlighter sources at laser facilities~\cite{MacDonald_PoP_2021}). Moreover, at laser facilities the capsule is typically large (several 100~$\mu$m wide), which provides an additional source of broadening.
To demonstrate the wider applicability of the ratios method, we consider a typical XRTS experimental setup at the NIF -- a Zn He-$\alpha$ backlighter probing a spherical capsule, and the spectra measured using the MACS spectrometer.

This setup presents a number of potential challenges for the method. First, the SIF broadening is substantially larger, which will require a much larger integration range to remove the SIF broadening than before. However, the MACS spectrometer has a very large spectral range of $>3$~keV, which will provide more than sufficient integration range to observe the convergence of the inferred temperature.

Second, and more problematic, is that the SIF varies substantially across the spectrometer, as shown in Fig.~\ref{fig:NIF}~(a). This is due to the out-of-focus geometry of the MACS spectrometer, which results in photons spread across conical arcs centred around the focused energy of 8.6~keV. Evidently this raises questions as to the applicability of the convolution approximation in Eqs.~(\ref{eq:XRTS_Convolution}) and~(\ref{eq:Ratios}). Nevertheless, this approximation routinely used in analysis~\cite{Tilo_Nature_2023,Dornheim_2025_Unraveling}, and we shall still assume it to hold here.

Third is that the large capsule size, source profile, and finite-size of the spectrometer would mean XRTS spectra over a range of scattering vectors would be measured on the detector. Moreover, conditions within the capsule are not uniform, with substantial variations in temperature and density predicted by radiative hydrodynamics modelling~\cite{Tilo_Nature_2023}. However, such considerations fall outside the scope of the current work, so we treat conditions within the capsule to be uniform. Still, these effects should be examined in detail in future works.

For the DSF, we model laser-compressed Be using the conditions reported in Row 1 of Table 1 in Ref.~\cite{Tilo_Nature_2023}; a mass density of 6.7~g~cm$^{-3}$, ionization state of 3, and a temperature of 110~eV. The modelling remains the same as before, except that the Stewart-Pyatt model~\cite{stewart1966lowering} is used to model the effect of ionization potential depression on the bound-free energy. The DSFs are calculated for an incident beam energy of 9.0~keV and scattering angles of 30$^\circ$, 75$^\circ$, and 120$^\circ$, which are available using the Gbar experimental platform at the NIF.

The backlighter used in Ref.~\cite{Tilo_Nature_2023} consists of He-$\alpha$ lines from laser-heated Zn, without filters. To model this profile, we use the FLYCHK code~\cite{FLYCHK} to simulate the Zn emission at $T = 4$~keV and an electron density $n_e = 1\times10^{21}$~cm$^{-3}$, as is done in Ref.~\cite{MacDonald_PoP_2021}.
Radiography measurements in Ref.~\cite{Tilo_Nature_2023} give the capsule diameter as 200~$\mu$m. To model the source in the ray tracing simulations, we sample a uniform sphere of the same diameter centred at the origin. For the non-dispersive region of interest, we integrate of 500 pixels. Finally, the simulations are run using $1\times10^8$ photons sent to the crystal.

The ray traced XRTS spectra are shown in Fig.~\ref{fig:NIF}~(b). A ray traced simulation of the elastic feature is also plotted to provide an indication of the shape and width of the SIF broadening. The FWHM of the feature is 91~eV, which is substantially larger than the $\sim 3$~eV broadening seen in the previous simulations.

The convergence of the inferred temperature for the three different ITCF ratios is plotted in Fig.~\ref{fig:NIF}~(c). 
As expected, the onset in convergence is much slower than for the previous experimental setup considered, requiring an integration limit $x \sim 400$--600~eV, depending on the ratio.
The three different inferred temperatures are in good agreement with each other, but overestimate it to be 114--116~eV compared to the true value of 110~eV. That the temperature is reasonably close to the true value is perhaps surprising given the substantial variation in reflectivity across the spectral range; see Fig.~\ref{fig:NIF}~(a).
Overall, this is a promising result that even very complicated instrument functions can be effectively accounted for within the ratios method. Additionally, convergence in the temperature is observed comfortably within the spectrometer's spectral range.

\section{Summary and Discussion\label{sec:summary}}

We have presented and investigated a new method for the extraction of temperature directly from XRTS spectra, without the need for a model of the dynamic structure factor nor for the source-and-instrument function of the setup.
Recent work has suggested that temperatures can be directly extracted from XRTS spectra, without the need for models of the DSF~\cite{Dornheim_T_2022,Dornheim_T2_2022}, using the convolution theorem for the two-sided Laplace transform to deconvolve the SIF from the spectrum -- here, we have verified that, provided the SIF measured at the elastic peak is well-characterised, the Laplace deconvolution can be used to very accurately extract temperature, despite the fact that the SIF does not strictly broaden the DSF as a convolution.
However, owing to the difficulty in accurately determining the crystal response function, the SIFs used so far in application of the ITCF method to experiment have been models, which has reintroduced model-dependency to the method.

Therefore, we have demonstrated via ray tracing simulations of HAPG von H\'amos spectrometers that, provided the instrument functions of the two spectrometers are sufficiently similar, the deconvolution can in effect be carried out by taking the ratio of two Laplace-transformed XRTS spectra collected simultaneously.
This bypasses needing to explicitly include the SIF in the deconvolution process, and enables a truly model-free interpretation of temperature from XRTS spectra: a model is not needed for the DSF, nor for the SIF.

Under the assumption that the spectrometers have identical crystals, we have investigated the resilience of the ratio method to spectral noise and spectrometer misalignment. With regard to the former, we find that the lower temperatures require increasingly stringent requirements on the photon statistics since sufficient upshifted scattering signal over the quasi-elastic feature needs to be measured to probe detailed balance. The photon statistics also directly impact the ability to measure non-equilibrium effects via discrepancies in the inferred temperatures: lower photon counts introduce larger uncertainties in the extracted temperature, and therefore would require larger temperature discrepancies to confidently infer non-equilibrium effects.
We also find that the spectrometers do not need to be perfectly aligned to use the method, and that the allowable misalignment is largely dictated by the relative intensity of the elastic scattering compared to the inelastic scattering. If the elastic feature is relatively weak, large misalignments of even up to 10~mm can still yield accurate results. However, a very prominent elastic feature means good spectrometer alignment (to within $\sim 1$~mm) is required.

We also investigated the reliability of the method when the crystals' physical properties are different, namely for the mosaicity and the intrinsic rocking curve widths.
We find that there is some leeway in the differences between crystals, so they do not need to be identical. For mosaic crystals, this is very promising since their random nature means no two crystals are identical.
As with the spectrometer misalignment, the degree to which the crystals can have different mosaicities and IRC widths is largely dictated by the relative prominence of the elastic feature -- a weak elastic feature and strong inelastic scattering can still yield accurate results even when the crystals are very different. Conversely, if the elastic feature is very large, the crystals need to be quite similar to one another. There is some ability to optimise the setup by choosing which crystal should measure which scattering angle; e.g. whether to use the higher mosaicity crystal to measure to the higher or lower scattering angle. In general, some planning is required to determine the optimal geometry to best utilise the ratios method.

Additionally, we have examined the applicability of the method to broad SIFs that are typical at laser facility experiments, such as the NIF. The integration range required to observe the temperature converge is large, but comfortably lies within the spectrometer spectral range. The specific spectrometer considered in those simulations was the MACS spectrometer, whose instrument function varies substantially within the considered spectral range, and so is not a simple convolution. Despite this, the ratios method is able to extract reasonable temperatures from the simulated spectra.

The primary benefit of mosaic crystals is their very reflectivities, with mosaic graphite being the crystal with the highest known integrated reflectivity.
One alternative to get better consistency in the crystals' response functions would be to use perfect bent crystals in von H\'amos geometry (see e.g. Ref.~\cite{Zastrau_2014_Bent}). The rocking curve of the crystal is then very well-described by the Takagi-Taupin equations~\cite{Takagi_1962_Dynamical,Taupin_1964_Theorie}. The expected lower reflectivities of such crystals can be compensated for at XFEL facilities by collecting more frames at the desired conditions, though this may not be possible in cases of time and target availability constraints.

In practice, we expect that some measurement of the spectrometers' instrument functions will be useful, in particular for characterising the similarity of their mosaicities and IRC widths, as well as identifying at least the first drop in intensity due to the crystals' edges being reached. Finding this first drop is important for spectrometer spatial alignment, and in general these measurements are important for determining whether the crystals are sufficiently similar to use the ratio method for the systems that will be probed. Such IF measurements also provide useful backups in case the crystals are simply too different to use the ITCF ratio method. In this case, measurements of the IF are very useful for informing accurate modelling of the SIF to use in the standard ITCF method.

The ratio method is restricted to experimental setups that are able to measure XRTS spectra at multiple scattering angles simultaneously. At XFELs, measuring XRTS spectra from two scattering angles is very standard. However, space constraints (e.g. from other diagnostics or laser optics) may make it challenging to fit an additional spectrometer to measure at three or more scattering angles simultaneously. Likewise, at some laser facilities such as the NIF, XRTS measurements from multiple scattering angles simultaneously is not a standard setup and would need to be developed further. These situations would also benefit from good SIF characterisation to use the standard Laplace method -- either to extract a temperature from a single XRTS spectrum, or to measure non-equilibrium effects by comparing temperatures from two scattering angles.

Nevertheless, we anticipate that at existing and future experimental setups, the ITCF ratio method will prove very useful for directly extracting system properties, without the need for any input models. Moreover, given the ratio of two ITCFs contains a wealth of additional information that is contained in the individual ITCFs -- for example, about the spectral normalisation via the f-sum rule~\cite{dornheim2024fsum}, comparative Rayleigh weights between scattering angles~\cite{Dornheim_2024_Wr}, and ratio of the static linear density response functions~\cite{schwalbe2025staticlineardensityresponse} -- further developments of the ratio method will allow for other system properties beyond the temperature to be extracted directly from XRTS spectra.

\section*{Data Availability Statement}
The simulation datasets supporting this work are available at the following doi: 10.14278/rodare.4061.

\section*{Author Declarations}
The authors have no conflicts to disclose.

\section*{Acknowledgments}
The authors thank Stephanie B. Hansen for useful discussions on this work.

This work was partially supported by the Center for Advanced Systems Understanding (CASUS), financed by Germany’s Federal Ministry of Education and Research (BMBF) and the Saxon state government out of the State budget approved by the Saxon State Parliament.
This work has received funding from the European Union's Just Transition Fund (JTF) within the project \textit{R\"ontgenlaser-Optimierung der Laserfusion} (ROLF), contract number 5086999001, co-financed by the Saxon state government out of the State budget approved by the Saxon State Parliament.
This work has received funding from the European Research Council (ERC) under the European Union’s Horizon 2022 research and innovation programme
(Grant agreement No. 101076233, ``PREXTREME''). 
Views and opinions expressed are however those of the authors only and do not necessarily reflect those of the European Union or the European Research Council Executive Agency. Neither the European Union nor the granting authority can be held responsible for them.
Tobias Dornheim gratefully acknowledges funding from the Deutsche Forschungsgemeinschaft (DFG) via project DO 2670/1-1.
A.K. and A.D.B. were partially supported by the U.S. Department of Energy Science Campaign 1. Sandia National Laboratories is a multi-mission laboratory managed and operated by National Technology \& Engineering Solutions of Sandia, LLC (NTESS), a wholly owned subsidiary of Honeywell International Inc., for the U.S. Department of Energy’s National Nuclear Security Administration (DOE/NNSA) under contract DE-NA0003525. This written work is co-authored by employees of NTESS. The employees, not NTESS, own the right, title and interest in and to the written work and are responsible for its contents. Any subjective views or opinions that might be expressed in the written work do not necessarily represent the views of the U.S. Government. The publisher acknowledges that the U.S. Government retains a non-exclusive, paid-up, irrevocable, world-wide license to publish or reproduce the published form of this written work or allow others to do so, for U.S. Government purposes. The DOE will provide public access to results of federally sponsored research in accordance with the DOE Public Access Plan.
The work of M.P.B. was performed under the auspices of the U.S. Department of Energy by Lawrence Livermore National Laboratory under Contract No. DE-AC52-07NA27344 and supported by Laboratory Directed Research and Development (LDRD) Grant No.~25-ERD-047.
The authors gratefully acknowledge the computing time granted by the Resource Allocation Board and provided on the supercomputer Emmy/Grete at NHR-Nord@G\"ottingen as part of the NHR infrastructure. The calculations for this research were conducted with computing resources under the project mvp00024.

\bibliography{bibliography}
\end{document}